\documentclass[sigconf]{acmart}

\usepackage[para]{footmisc}

\usepackage{amsmath}
\usepackage{natbib}
\usepackage{amsfonts}
\usepackage{graphicx}
\usepackage{hhline}
\usepackage{hyperref}
\usepackage{enumerate}
\usepackage{multirow}
\usepackage{array}
\usepackage[para]{footmisc}
\usepackage{adjustbox}
\usepackage{balance}
\usepackage{subcaption}
\usepackage[normalem]{ulem}
\usepackage[show]{chato-notes}

\newcommand{\xw}[1]{\textcolor{black}{#1}}
\newcommand{\zx}[1]{\textcolor{black}{#1}}
\newcommand{\io}[1]{\textcolor{black}{#1}}
\newcommand{\update}[1]{\textcolor{black}{#1}}

\usepackage{marginnote}
\newcommand{\pageenlarge}[1]{\enlargethispage{#1\baselineskip}}
\usepackage{subcaption}

\usepackage{algorithm}
\usepackage{algorithmicx}
\usepackage{algpseudocode}

\usepackage{soul}

\copyrightyear{2024}
\acmYear{2024}
\setcopyright{acmcopyright}
\acmConference[SIGIR '24]{Proceedings of the 47th International ACM SIGIR Conference on Research and Development in Information Retrieval}{July 14--18, 2024}{Washington D.C., USA)}
\acmBooktitle{Proceedings of the 47th International ACM SIGIR Conference on Research and Development in Information Retrieval (SIGIR '24), July 14--18, 2024, Washington D.C., USA}

\begin{document}

\title{A Directional Diffusion Graph Transformer for Recommendation}

\author{Zixuan Yi}
\email{z.yi.1@research.gla.ac.uk}
\affiliation{%
 \institution{University of Glasgow}
 \city{Glasgow}
 \state{Scotland}
 \country{United Kingdom}
}

\author{Xi Wang}
\email{xi.wang@sheffield.ac.uk}
\affiliation{%
 \institution{University of Sheffield}
 \city{Sheffield}
 \state{England}
 \country{United Kingdom}
}

\author{Iadh Ounis}
\email{iadh.ounis@glasgow.ac.uk}
\affiliation{%
 \institution{University of Glasgow}
 \city{Glasgow}
 \state{Scotland}
 \country{United Kingdom}
}

\fancyhead{}

\begin{abstract}
In real-world recommender systems, implicitly collected user feedback, while abundant, often \io{includes} noisy false-positive and false-negative \xw{interactions}.
\io{The possible} \xw{misinterpretations of \io{the} user-item} interactions pose a significant challenge for traditional graph \zx{neural recommenders}. \xw{These approaches} 
aggregate \io{the} users' or items' neighbours \xw{based on} implicit user-item interactions \io{in order} to accurately capture \io{the users'} profiles.
\io{To account for and model possible noise in the users'} interactions in \zx{graph neural recommenders}, we \io{propose} a novel \textbf{Diff}usion \textbf{G}raph \textbf{T}ransformer (\textbf{DiffGT}) model for top-$k$ recommendation.
Our DiffGT model employs \io{a} diffusion process, which includes a forward phase for gradually introducing noise to implicit interactions, followed by a reverse process to iteratively refine the representations of \io{the} users' hidden preferences (i.e., \io{a} denoising \io{process}).
In our \io{proposed} approach, given the inherent anisotropic structure observed in \io{the} user-item interaction graph,
\xw{we specifically use anisotropic and directional Gaussian noises in the forward diffusion process. \io{Our} approach differs from the \io{sole} use of isotropic Gaussian noises in \io{existing} diffusion models.} 
\xw{In the reverse diffusion process,
to reverse the effect of noise added earlier and recover the true \io{users'} preferences,
we integrate a graph transformer architecture with a linear attention module to denoise the noisy user/item embeddings in an effective and efficient manner. In addition, such a reverse diffusion process is further guided by personalised information (e.g., interacted items) to enable \io{the} accurate estimation of \io{the users'} preferences on items.}
Our extensive experiments conclusively demonstrate the superiority of our proposed graph diffusion model over ten existing state-of-the-art approaches across three benchmark datasets.
\pageenlarge{3}

\end{abstract}




\maketitle

\section{Introduction}\label{sec:intro}

\begin{figure}[tb]
    \begin{subfigure}[t]{0.41\linewidth}
        \includegraphics[trim={4cm 0cm 4cm 0cm},clip,width=1\linewidth]{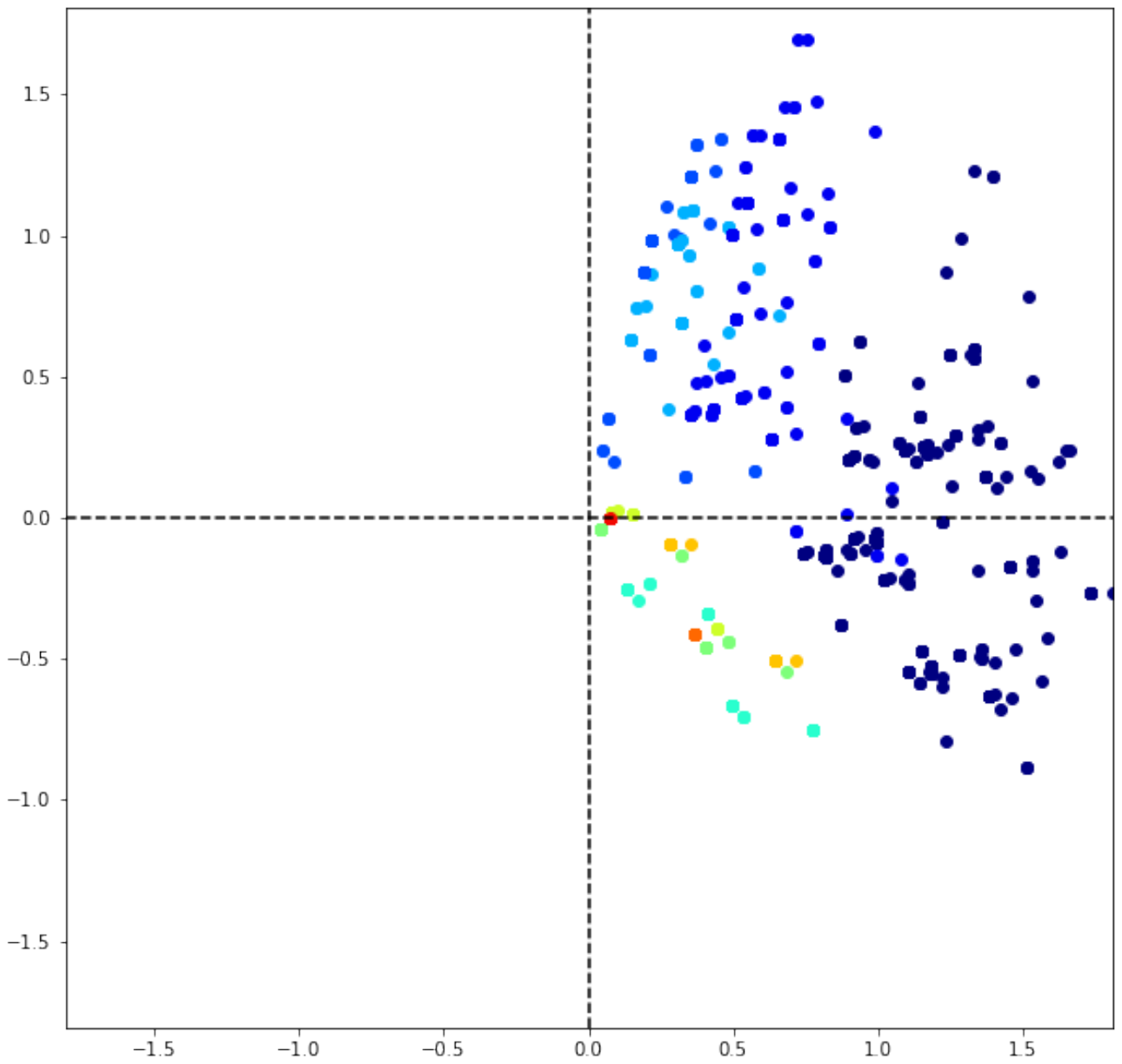}
        \captionsetup{labelformat=empty}
        \caption{(a) MovieLens-1M}
        \label{fig:subfig1}
    \end{subfigure}
    \begin{subfigure}[t]{0.41\linewidth}
        \includegraphics[trim={0cm 4.6cm 0cm 4cm},clip,width=1\linewidth]{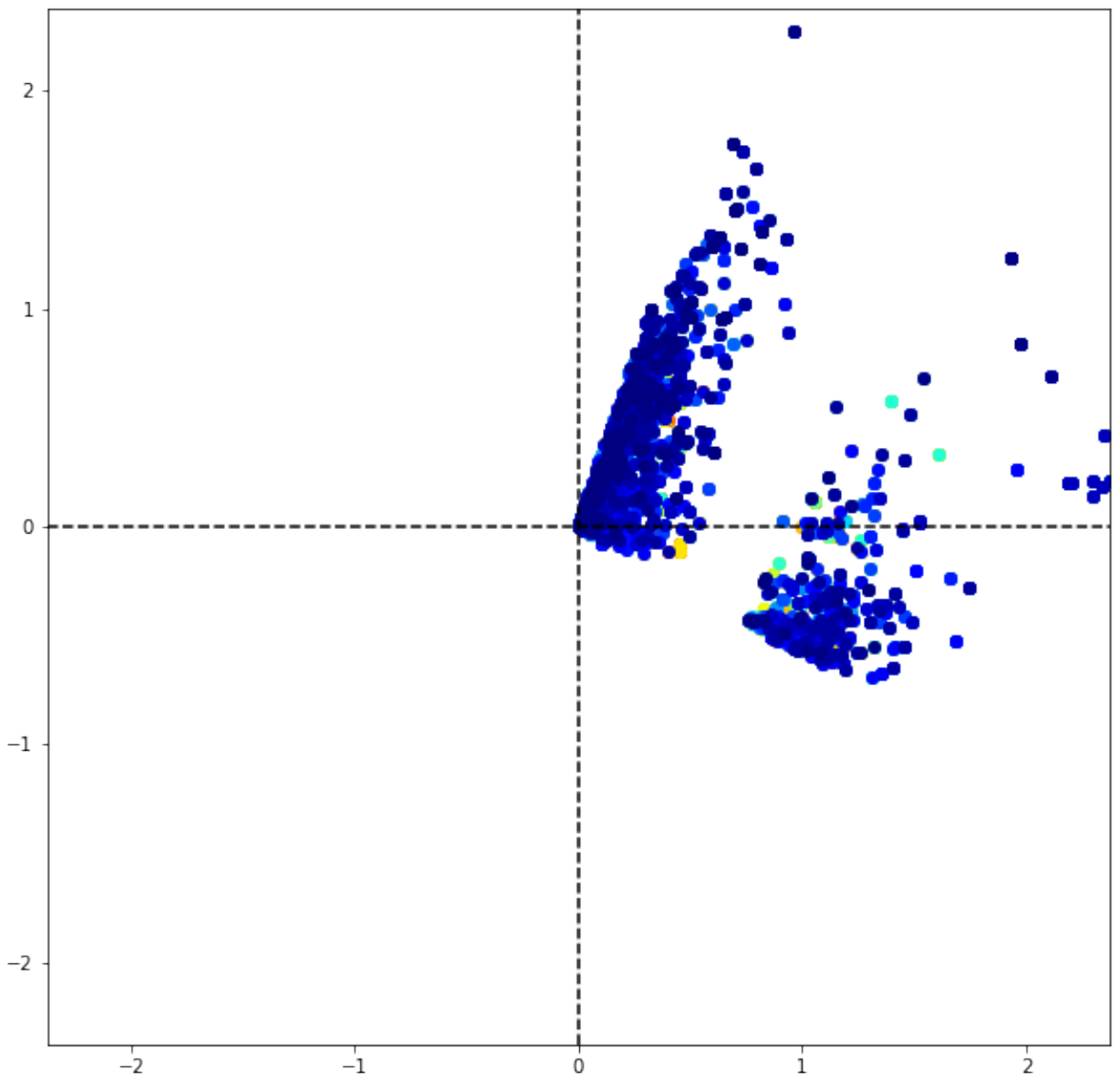}
        \captionsetup{labelformat=empty}
        \caption{(b) Foursquare}
        \label{fig:subfig2}
    \end{subfigure}
    \captionsetup{labelformat=simple}
    \caption{2D visualisation of the data using SVD decomposition. (a) Visualisation of the item
representations in the MovieLens-1M dataset, with different colours indicating the genres of movies. (b) Visualisation of all tags of venues in the Foursquare dataset.}
\label{fig:svd}
\vspace{-6mm}
\end{figure}

In recommender systems, collaborative filtering stands as an effective technique, primarily relying on implicit interactions~\cite{jawaheer2014modeling}. These implicit interactions encompass user-item interactions such as clicks, browsing history, and bookmarks, which are \io{typically} abundant in collaborative \io{filtering setups}. 
Unlike explicit interactions, \xw{where users provide intentional} 
input like ratings or reviews, implicit interactions represent \io{the users'} preferences in a one-dimensional, positive-only format~\cite{wu2016collaborative}. \xw{However, they often encompass noisy interactions, making interpretation challenging.}
For example, 
\xw{a} high volume of clicks 
\xw{may not necessarily indicate \io{an} actual intent to purchase \io{items}. Similarly, a purchase might not always reflect satisfaction, as indicated by instances where purchases are followed by negative reviews.}
\xw{These discrepancies can frequently be attributed to \io{biases in the presentation of} items, such as the position bias, which considerably influences \io{the} users' initial impressions and \io{their} subsequent interactions \io{with the items}~\cite{wang2021denoising}.}
\xw{This notable gap between \io{the user} interactions and \io{the} actual user satisfaction \io{raises} a critical challenge: implicit interactions, while rich in collaborative information, often fail to accurately represent the full spectrum of \io{the users'} preferences due to the prevalence of noisy data. Consequently, the task of uncovering \io{the} users' true preferences amidst this inherent noise becomes crucial. }

\pageenlarge{3}
\looseness -1 \io{Several} approaches have been used to mine more effective collaborative signals from implicit interactions~\cite{wu2016collaborative, wang2021denoising, liang2018variational, wang2023diffusion},
\zx{thereby aiming to enhance the performance of \io{the} top-$k$ recommendation task}. 
\io{Some prior} studies~\cite{shenbin2020recvae,wu2016collaborative,liang2018variational} \io{leveraged} Auto-Encoders (AEs) 
to denoise the noisy implicit user-item interactions by reconstructing \io{the} user/item \io{embeddings} with point-wise distance losses.
For example, 
CDAE~\cite{wu2016collaborative} \io{introduced} random noises to \io{the} users’ interactions during training and \io{employed} an auto-encoder for denoising. 
MultiVAE~\cite{liang2018variational} \io{used} variational auto-encoders to model implicit \io{interactions} between \io{the} users and items with multinomial distribution and \io{optimised} the evidence lower bound loss between the distribution of \io{the} encoder and decoder's output.
The recent DiffRec~\cite{wang2023diffusion} also \io{leveraged} a diffusion model with an autoencoder architecture to denoise the implicit interactions. 
Conceptually, diffusion models gradually corrupt the user-item interactions in a tractable forward process and learn the reverse reconstruction iteratively \xw{in the direction of recovering the \update{original} interactions}~\cite{wang2023diffusion}.
In particular, DiffRec
\io{leveraged} continuous diffusion at the embedding level,
gradually \io{adding} normal Gaussian noises -- a type of \zx{\textit{isotropic noise}}~\cite{yang2023directional} -- into the encoded user/item embeddings as noisy latent variables. Subsequently, it
\io{refined} the obtained noisy embeddings in a denoising manner. 
To understand the implications of the used noise, we conduct a preliminary analysis \xw{on two commonly \io{used}} recommendation \xw{datasets} using 2D Singular Value Decomposition (SVD).
Figure 1(a) shows data instances in various colours representing each item as per \io{the} movie genre distribution in the MovieLens-1M dataset, while Figure 1(b) represents the item distribution according to \io{the} venue tags \io{(examples of tags include "brunch" and "bar")}  
in the Foursquare dataset.
For example, the projected data from the Foursquare dataset exhibit strong anisotropic structures along only a few directions rather than spreading over the distribution space.
This observation indicates that the recommendation data inherently features distinctive anisotropic and directional characteristics, which may not align with the assumptions underlying the use of isotropic Gaussian noise. 
Consequently, the integration of isotropic noise in existing diffusion-based recommendation models may lead to 
\zx{less distinguishable}
user/item embeddings, thereby complicating \io{the preservation of} item heterogeneity \zx{by obscuring the unique attributes of items.}
Therefore, \io{we argue that} the use of \zx{\textit{directional noise}} is more suitable, as it aligns with the anisotropic structures typically observed in the recommendation data.
The adoption of directional noise not only addresses the issue of distinguishable embeddings but also aligns with the underlying data patterns, thereby enhancing the model's ability to accurately capture the unique characteristics of each user/item.

\pageenlarge{2}
\looseness -1 Inspired by the remarkable success of diffusion models in different domains, such as image synthesis~\cite{rombach2022high,dhariwal2021diffusion}, sequential recommendation~\cite{liu2023diffusion,li2023diffurec,wang2023diffusion} and graph/node classification tasks~\cite{gasteiger2019diffusion,kong2023autoregressive,vignac2023digress,yi2022multi}, \io{in this paper, we propose an effective} diffusion model from the inherent characteristics of recommendation data. \xw{Specifically, due to the \io{effective} performance of graph neural network models in modelling user-item interactions~\cite{he2020lightgcn,wang2019neural,yu2022self} as well as \io{the} wide application of diffusion to graph models}~\cite{feng2023diffuser,wu2023difformer,hua2023mudiff}, we \io{focus} on 
\xw{validating our novel diffusion approach when applied} to \zx{graph neural \io{recommendation systems}}. 
However, the common practice of implementing diffusion \io{in} \zx{graph neural models} is based on discrete diffusion, which necessitates the use of a transition matrix to update the binary states (true/false) of interactions at
each diffusion step~\cite{chen2023d4explainer}.
This process updates the edges' state\zx{s} in the adjacency matrix according to the transition probabilities in the matrix, \io{increasing} the computational complexity of the model.
This discrete diffusion approach, while tailored to discrete data such as graph data, has limitations in terms of generalisation. 
\zx{Despite this limitation}, \zx{graph neural recommenders} are notably effective in exploiting the interactions between the nodes, 
\zx{which are the key features in modelling collaborative signals} in recommender systems.
\zx{Considering these aspects and the observed directional structure in the recommendation data (as shown in Figure~\ref{fig:svd}), we aim to determine (1) the \zx{most effective} diffusion methods (discrete vs. continuous) and (2) the \zx{most effective} type of noise (normal vs. directional) in recommender systems.}
\io{We introduce a \io{new} diffusion graph transformer (DiffGT) model, which \xw{\zx{aims for both an} effective and efficient integration of \io{a} diffusion process \io{tailored} to recommender systems}}. \io{Our model}  
incorporates directional noise and a linear transformer on the user/item node embedding level. In addition, to show the \zx{generalisation} of our proposed approach, we extend its application to other recommendation \zx{models}, 
\xw{including} knowledge graph\xw{-augmented} recommendation \zx{models} and sequential recommendation \zx{models}.


\io{Our contributions in this paper} can be summarised as follows:
(1) We propose a \zx{new} graph transformer model with a diffusion process for top-$k$ recommendation.
To the best of our knowledge, we are the first to apply diffusion to denoise implicit interactions in \zx{recommender systems}.
(2) We introduce the anisotropic and directional noise in the forward process of diffusion for an improved performance, which is \zx{different} from the existing use of normal isotropic noise.
(3) We implement the diffusion process on the \io{embeddings} level and leverage a linear transformer to denoise the obtained noisy embeddings for efficient purposes. \zx{In addition, we sample fewer steps in the reverse process for more efficient denoising.}
(4) We also experiment on side information-enhanced user-item interaction graphs and conclude that our \zx{proposed} graph diffusion transformer can better mine effective user/item embeddings from more noisy interaction graphs \zx{compared to} existing \zx{recent} baselines.
(5) We extend the use of the directional noise in the diffusion process, along with a linear transformer, to \zx{address} both knowledge graph recommenders and sequential recommenders. 
Our findings show that 
\zx{the use of the directional diffusion and the linear transformer}
demonstrate \io{an} effective generalisation \zx{to both knowledge graph-enriched and sequential recommendation models.}

\pageenlarge{2}
\section{Related Work}\label{sec:related}
In this section, we present an overview of related methods that are relevant to our work, namely denoising for recommendation and \zx{graph neural recommenders}. \xw{\io{We also position}} our work with respect to the \io{existing} literature.

\noindent \textbf{Denoising in Recommendation:}
\looseness -1 In recommender systems, the presence of noise poses significant challenges to implicit interactions.
\io{The noise} can skew the understanding of \io{the users'} preferences and obscure \io{their} true behavioural patterns, complicating the \xw{accurate} prediction of \io{the users'} interests~\cite{zhang2023sled}.
The challenge \xw{and the necessity} of mitigating noise in implicit user-item interactions have led to the exploration of various denoising techniques \xw{for} recommender systems. 
Traditional approaches~\cite{wu2016collaborative,walker2022recommendation,wang2023diffusion,li2023diffurec,ye2023towards,yi2023contrastive,yi2023graph} have primarily focused on using advanced neural models.
Auto-encoder-based approaches like CDAE~\cite{wu2016collaborative} corrupted the implicit interactions of users with random noises and then reconstruct the original input with auto-encoders during training. 
Recent diffusion-based recommendation models~\cite{walker2022recommendation,wang2023diffusion,li2023diffurec} principally \io{enabled} the used auto-encoder architecture to effectively \io{denoise} implicit interactions at 
different noise levels and remove the noise progressively in a step-wise manner.
For example, 
DiffRec~\cite{wang2023diffusion} \io{applied} the isotropic Gaussian noise on \io{the} user's interaction embeddings and \io{used} a shared Multilayer Perceptron (MLP) for multi-step prediction, while maintaining a step-by-step approach to predict the subsequent state \zx{of these user interaction embeddings.}
However, these denoising methods assume \io{that} the embeddings are uniformly distributed and incorporate isotropic Gaussian noise into the user/item embeddings~\cite{ortiz2020neural}.
\xw{Instead of employing} isotropic noise, our model \io{uses} anisotropic and directional noise in the diffusion process to allow \zx{a more}
effective \zx{and tailored} interaction prediction from the noisy implicit interactions.
\io{We argue that the} choice of anisotropic and directional noise \io{better aligns} with the inherent anisotropic structures 
observed in Figure~\ref{fig:svd}.
To the best of our knowledge, \io{our present work in this paper is the} first to incorporate anisotropic and directional noise in a diffusion recommender.
Furthermore, {the} existing {diffusion recommender (i.e., DiffRec)} essentially \zx{elaborate{s}} an unconditional generation paradigm.
\zx{Hence,} we integrate personalised information (e.g. users' interacted items) as a condition to guide the diffusion process, enabling more effective user-item interaction predictions that \zx{better} aligns with user interests.

\pageenlarge{3}
\noindent \textbf{Graph Neural Recommenders:}
\looseness -1 The essential principle of a \zx{graph neural recommender} is to iteratively aggregate the features of neighbouring nodes into a representation of the target node~\cite{kipf2016semi}.
However, the user-item graph includes implicit \io{feedback} as edges, \xw{which has a high chance of adding} false-negative interactions between \io{the} users and items. 
These noisy edges \io{can} impede the \zx{graph neural recommenders} \io{from accurately modelling the} user/item \io{representations} \xw{after} aggregating inherent noises within the user-item interaction graph~\cite{jiang2023adaptive}.   
Recent graph studies~\cite{wang2019neural, he2020lightgcn} have found that \xw{applying} \io{an} appropriate dropout of \xw{included} edges from the interaction graph \xw{can result} in a performance improvement.
Specifically, both LightGCN~\cite{he2020lightgcn} and SGL~\cite{he2020lightgcn} employed edge dropout before obtaining \io{the} user/item embeddings from their neighbours while SGL extended this approach by employing contrastive learning to generate and differentiate two different \zx{user/item embeddings} created by the dropout \zx{operations}.
GFormer further \zx{used} contrastive learning \zx{for} dropping edges by re-weighting the importance of the edges through a separate transformer model and contrasting two obtained user/item embeddings.
Despite the success of the \zx{graph neural recommenders}, the existing \zx{graph neural recommenders} do not explicitly \io{denoise} the user-item interactions from the user's \io{feedback}. 
\zx{Considering this gap,}
to address the noisy interactions in the user-item graph, we devise a \zx{new} diffusion graph transformer (DiffGT) model to denoise the implicit interactions among users and items.
Specifically, the incorporation of the transformer architecture into the graph data structure has been proposed as a potential solution~\cite{feng2023diffuser,wu2023difformer,hua2023mudiff,yi2023large,yi2023precontrastive}, analogously mirroring the difficulty of denoising the \io{non-relevant} information and ambiguities in natural language processing~\cite{zou2023diffusion}.
However, the use of a graph transformer architecture for denoising within the diffusion process in the recommendation scenario has not \xw{yet been} investigated in the literature.
Note that our DiffGT model differs from GFormer, which only uses a separate transformer to construct \zx{another} view of user/item embeddings.
\zx{In contrast, our approach \io{uses} a cascaded graph transformer architecture where a transformer is paired with a graph encoder to denoise the noisy user/item embeddings in a diffusion process.}
To the best of our knowledge, our DiffGT model is the first {graph neural} approach to incorporate a transformer to mitigate noisy implicit interactions in recommender systems.
\zx{We} also extend the application of our diffusion graph transformer designs to other recommendation contexts, including knowledge graph recommendation and sequential recommendation tasks.




\section{A Directional Diffusion Model}
In this section, we start by introducing \io{some} preliminary \io{background information} \io{as well as} the notations \io{we will be using in the remainder of the paper}.
Then, we identify the limitations of incorporating the diffusion process into \zx{graph neural recommenders} \xw{-- the \io{models we initially focus on} among recommendation techniques --} and introduce our proposed DiffGT model.
Subsequently, we \io{describe} the injection of directional noise within the diffusion process \io{into} our \io{proposed} \zx{Diffusion Graph Transformer (DiffGT) model} 
\xw{while tailoring each component to the forward and reverse diffusion processes.} 
\xw{Next, we also \io{present} the optimisation strategy of our DiffGT model \io{for the} top-$k$ recommendation \io{task}.}

\pageenlarge{3}
\subsection{Preliminaries}\label{sec:pre}
A vanilla diffusion model \xw{involves two major components}: the forward and reverse processes, \xw{which} \io{ use} latent variable modelling to enable the progressive generation of refined representations~\cite{sohl2015deep}.

\noindent \textbf{Forward Process:}
In the forward process of the diffusion model, a data point initially sampled from a real-world distribution, $\mathrm{x}_0 \sim q(\mathbf{x})$, is progressively corrupted through a Markov chain into \io{a} standard Gaussian noise, represented as $\mathbf{x}_T \sim \mathcal{N}(0, \mathrm{I})$. 
For each forward step $t \in[1,2, \ldots, T]$,
\zx{this corruption transforms the original representation} 
$\mathbf{x}_0$ 
into \io{a} noisy representation $\mathbf{x}_t$ at each step $t$: 
\begin{equation}\label{eqn:forward}
\begin{aligned}
    q\left(\mathbf{x}_t \mid \mathbf{x}_{t-1}\right) & = \mathcal{N}\left(\mathbf{x}_t; \sqrt{1-\beta_t} \mathbf{x}_{t-1}, \beta_t \mathbf{I}\right), \\
    & = \sqrt{1-\beta_t} \mathbf{x}_{t-1} + \sqrt{\beta_t} \epsilon, \quad \epsilon \sim \mathcal{N}(0, \mathrm{I})
\end{aligned}
\end{equation}
where $\mathcal{N}$
is a Gaussian distribution, 
and $\beta_t \in(0,1)$ \xw{controls the level of \io{the} added noise at step $t$}. 
This \zx{approach} \xw{shows the flexibility of \io{the} direct sampling of} $\mathbf{x}_t$ conditioned on the input $\mathbf{x}_{t-1}$ at an arbitrary diffusion step $t$ 
from a random Gaussian noise $\epsilon$.


\noindent \textbf{Reverse Process:}
\looseness -1 The primary purpose of \io{the} diffusion models is to learn a denoising model capable of \xw{removing the added noise to the data and \io{to} gradually \io{recover} the initial distribution.}
\io{Indeed, once} \xw{a} noisy embedding $\mathbf{x}_{t}$ is obtained, 
the reverse process aims to denoise this $\mathbf{x}_{t}$, 
with a trajectory towards the direction of $\mathbf{x}_0$ and \io{to} gradually 
\xw{recover the initial} representation $\mathbf{x}_0$.
The transition $(\mathbf{x}_t \rightarrow \mathbf{x}_{t-1} \rightarrow$ $\left., \ldots, \rightarrow \mathbf{x}_0\right)$ is defined as follows:
\begin{equation}\label{eqn:reverse}
\begin{aligned}
p\left(\mathbf{x}_{t-1} \mid \mathbf{x}_t, \mathbf{x}_0\right) & =\mathcal{N}\left(\mathbf{x}_{t-1} ; {\mu_\theta}_t\left(\mathbf{x}_t, \mathbf{x}_0\right), {\beta}_t \mathbf{I}\right) \\
{\mu_\theta}_t\left(\mathbf{x}_t, \mathbf{x}_0\right) & =\frac{\sqrt{\bar{\alpha}_{t-1}} \beta_t}{1-\bar{\alpha}_t} \mathbf{x}_0+\frac{\sqrt{\alpha_t}\left(1-\bar{\alpha}_{t-1}\right)}{1-\bar{\alpha}_t} \mathbf{x}_t \\
{\beta}_t & =\frac{1-\bar{\alpha}_{t-1}}{1-\bar{\alpha}_t} \beta_t
\end{aligned}
\end{equation}
where $\mu$ and $\sigma^2$ are \io{the} mean and variance \zx{of the user/item embeddings}, respectively; $\alpha_t=1-\beta_t$ \io{while} $\bar{\alpha}_t=\prod_{t^{\prime}=1}^t \alpha_{t}$. 
The learning of the mean $\mu_\theta$ is based on a neural network $f_\theta$ parameterised by $\theta$. 
As such, to effectively \xw{recover} $\mathbf{x}_0$, the use of neural models, such as \io{a}  Transformer~\cite{vaswani2017attention} or \io{a} U-Net~\cite{dhariwal2021diffusion}, is commonly used in practice~\cite{ho2020denoising}.

\noindent \textbf{Optimisation:}
As described above, the diffusion model parameterises both the mean $\mu_{\theta}$ and variance $\sigma_{\theta}$ and learns to approximate the real data distribution during the reverse process. The model parameter $\theta$ is updated by optimising the  stabilised and simplified loss function as follows:
\begin{equation}\label{eqn:diffusion}
\begin{aligned}
\mathcal{L}_{\text{diffusion}}=\left\|\epsilon-\epsilon_\theta\left(\sqrt{\bar{\alpha}_t} x_0+\sqrt{1-\bar{\alpha}_t} \epsilon, t\right)\right\|^2
\end{aligned}
\end{equation}
where $\epsilon_\theta(\cdot)$ 
denotes the function \update{that can be instantiated by a deep neural network to approximate $\epsilon$.} 
Note that this loss function is simplified from a variational lower-bound loss to prevent \io{an} unstable model training~\cite{wang2023diffusion,li2023diffurec}.

\looseness -1 Following the preliminaries of the diffusion process, its application in recommendation models presents two challenges,
primarily due to the following constraints: \textbf{\textit{Limitation 1:} Anisotropic and directional data structures:} As illustrated in Section~\ref{sec:intro}, \io{the} recommendation \io{data} often exhibits unique anisotropic and directional structures. 
The prevalent application of \io{a} normal Gaussian noise assumes the uniform distribution of data in the recommendation task. \zx{This assumption \io{does not} account for the unique and directional nature of \io{the} recommendation data, thus impeding } \io{the} effective \zx{learning} of the user/item representation\zx{s}.
\textbf{\textit{Limitation 2:} Uncondition\zx{ed} diffusion paradigm:} \io{The current} diffusion \io{approaches} in \io{recommendation} essentially \zx{perform} an uncondition\zx{ed} generation paradigm~\cite{wang2023diffusion}.
\zx{This approach \zx{presents} a limitation in effectively denoising the noisy user/item embeddings in recommender systems.}
\zx{Consequently,} there is a pressing need to \zx{refine these diffusion approaches by conditioning the diffusion process.}

\pageenlarge{2}
\subsection{Top-$k$ Recommendation Task}
\looseness -1 Let $\mathcal{U}$ and $\mathcal{I}$ denote the user and item set, respectively. \io{In the case of a graph neural network-based recommendation model,}
we \io{use} an interaction bipartite graph $\mathcal{G}$ for both users $\mathcal{U}$ and items $\mathcal{I}$, where nodes represent users/items and edges indicate user-item interactions. 
$N=|\mathcal{U}|+|\mathcal{I}|$ is the \io{number of nodes}, $\mathbf{A} \in \mathbb{R}^{N \times N}$ is the adjacency matrix, and $\mathbf{X}=\left(x_1, x_2, \cdots, x_N\right)^{\mathrm{T}} \in \mathbb{R}^{N \times d}$ is the matrix \zx{of the node features}.
In this paper, we aim to estimate \io{the users'} preferences \xw{through a neural recommender $f_{\theta}$ \zx{(e.g., a graph transformer model)}}, \io{which} can recommend the top-$k$ \io{items} for a target user $u$.

\subsection{Directional Gaussian Noise}\label{sec:directional}
As \zx{outlined in \textbf{\textit{Limitation 1}}}, our \xw{preliminary \io{analysis}} uncovered a pivotal factor contributing to the underwhelming performance of \zx{out-of-the-box} diffusion models \zx{in the top-$k$ recommendation task}: the \zx{observed} 
directional structure of \io{the} data. 
This further \io{highlights} the significance of investigating and tackling the challenges posed by anisotropic structures within diffusion models in the recommendation task.
To tackle this challenge, we \zx{inject} the directional noise in the forward diffusion process.
\zx{The transformation \update{from the isotropic to the anisotropic Gaussian noise ($\epsilon \rightarrow \epsilon^{\prime}$)} ensures a better alignment with the inherent directional structures present in \io{the} recommendation data.}
As a result, the user/item node feature $x_{t, i} \in \mathbf{R}^d$ of node $i$ at time step $t$ is obtained as follows:
\begin{equation}\label{eqn:noise1}
\begin{aligned}
x_{t} & =\sqrt{\bar{\alpha}_t} x_{0}+\sqrt{1-\bar{\alpha}_t} \epsilon^{\prime}, \\
\end{aligned}
\end{equation}
\begin{equation}\label{eqn:noise2}
\begin{aligned}
\epsilon^{\prime} & =\operatorname{sgn}\left(x_{0}\right) \odot|\bar{\epsilon}|, \\
\end{aligned}
\end{equation}
\begin{equation}\label{eqn:noise3}
\begin{aligned}
\bar{\epsilon} & =\mu+\sigma \odot \epsilon,
\end{aligned}
\end{equation}
where $x_{0}$ \io{holds} the raw \io{embeddings} of \zx{both users and items}. The symbol $\odot$ denotes the point-wise product. 
In Equation~\eqref{eqn:noise2}, we align the noise $\epsilon^{\prime}$ with the \zx{embeddings} $x_{0}$
to ensure that they share the same coordinate signs. This guarantees that adding noise does not result in noisy features pointing in the opposite direction of $x_{0}$. By preserving the directionality of the original \zx{embeddings}, \io{the} constraint \io{in~\zx{Equation}~\eqref{eqn:noise2}} plays a crucial role in maintaining the inherent data structure during the forward diffusion process. 
This, in turn, ensures the application of directional noise to avoid a fast decline in the signal-to-noise ratio,
preserving the essential information of the anisotropic structures~\cite{yang2023directional}.
In Equation~\eqref{eqn:noise3}, we transform the data-independent isotropic noise into an anisotropic noise.
By doing this, each coordinate of the noise vector shares the same empirical mean and \zx{the same} empirical standard deviation as the corresponding coordinate in the recommendation data. 
As such, the use of directional noise in our \io{proposed} DiffGT model \io{allows} to facilitate \io{an} effective user/item representation learning from the anisotropic structure of the latent space. 
\zx{In Section~\ref{ss:ablation}, we conduct experiments to ascertain the positive impact of directional noise.} \looseness -1

\begin{figure}[tb]\label{fig:arc}
        \captionsetup{labelformat=empty}
        \includegraphics[trim={1.5cm 7.5cm 2cm 8cm},clip,width=1\linewidth]{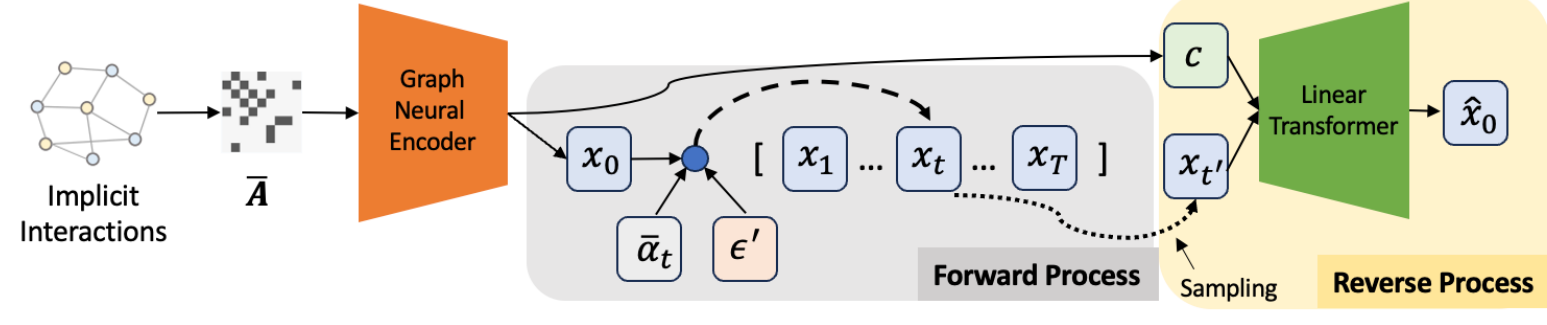}
        \captionsetup{labelformat=simple}
    \vspace{-4mm}
\caption{An illustration of our DiffGT architecture.}
\vspace{-4mm}
\end{figure}

\subsection{Diffusion Graph Transformer (DiffGT)}\label{sec:diffgt}

\begin{figure*}[tb]
\vspace{-10mm}
    \begin{subfigure}[t]{0.50\linewidth}
        \includegraphics[trim={4cm 5cm 4cm 5cm},clip,width=1\linewidth]{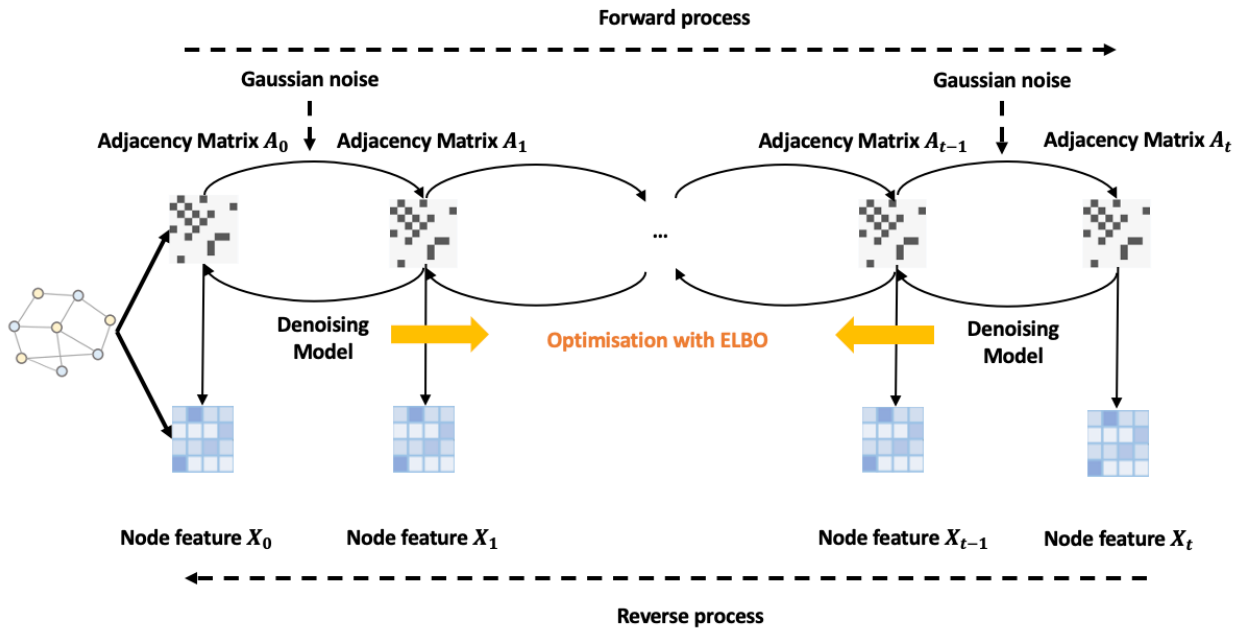}
        \captionsetup{labelformat=empty}
        \caption{Figure 3(a): An illustration of the discrete graph diffusion.}
    \end{subfigure}
    \begin{subfigure}[t]{0.49\linewidth}
        \includegraphics[trim={0cm 9.5cm 0cm 9.5cm},clip,width=1\linewidth]{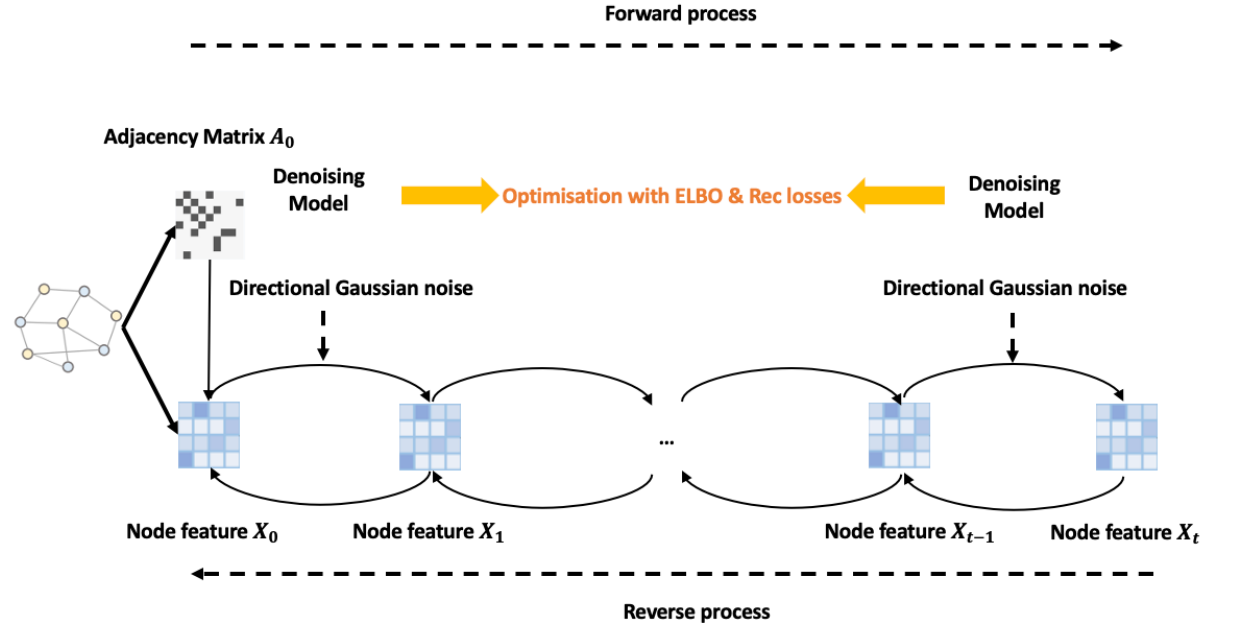}
        \captionsetup{labelformat=empty}
        \caption{Figure 3(b): An illustration of our continuous graph diffusion.}
    \end{subfigure}
    \captionsetup{labelformat=simple}
\label{fig}
\vspace{-4mm}
\end{figure*}
\looseness -1 Diffusion models have \io{led to} remarkable \io{successes} in various inverse problems in different domains, such as image synthesis~\cite{rombach2022high,dhariwal2021diffusion} and graph/node classification tasks~\cite{gasteiger2019diffusion,kong2023autoregressive,vignac2023digress}.
\io{As discussed in Section~\ref{sec:related}, }\xw{recent recommender systems have also adopted \zx{out-of-the-box} diffusion models for denoising \io{the} user embeddings. These models, however, overlook the anisotropic nature of recommendation data. \zx{In this paper, we} leverage \io{advanced user-item interaction modelling in} \zx{graph neural recommenders}~\cite{he2020lightgcn,yu2022self} 
to investigate the impact of applying directional Gaussian noise \zx{in the top-$k$ recommendation task.}
\io{Integrating} the diffusion process with \zx{graph neural recommenders} presents additional challenges:}
\textbf{\textit{Limitation 3} -- \io{Non-}informative bipartite interaction graph:} In contrast to diffusion models in the image synthesis task~\cite{rombach2022high}, which \io{start} with raw images as ground truth, the bipartite user-item interaction graphs in \io{the} recommender systems are inherently noisy and less informative, thus not \zx{necessarily} representing the `ground truth'. 
\textbf{\textit{Limitation 4} -- Insufficient model architecture:} 
The prevalent architectures for diffusion models in various domains~\cite{dhariwal2021diffusion,kong2023autoregressive,wang2023diffusion}, notably \zx{variational auto-encoders} and transformers, are not compatible with \zx{graph neural recommenders \zx{due to the lack of learning parameters and limited expressiveness~\cite{cai2023expressive}.}} 
Hence,  this mismatch necessitates a \update{graph-adapted architecture} to fully leverage the potential of diffusion methods in enhancing the top-$k$ recommendation. \looseness -1

\looseness -1 \zx{To address the above two limitations}, 
we introduce \io{the} DiffGT model, which
\zx{incorporates a novel graph transformer architecture to improve the diffusion \io{for an enhanced} top-$k$ recommendation performance.}
\zx{Additionally, t}his model more effectively denoises the implicit interactions by leveraging side information, thereby \zx{developing} a more accurate representation of a `ground truth' interaction graph.
\zx{As illustrated in Figure~2, our DiffGT architecture consists of two main components: a graph neural encoder and a linear transformer.}
\zx{Given \io{an} interaction graph, we obtain the adjacency matrix $\overline{\mathbf{A}}$, \zx{which is} enriched by the \zx{used} side information. For the forward process, once we obtain the original user/item embeddings $x_0$ encoded by the graph neural encoder, we gradually inject a scheduled directional noise $\epsilon^{\prime}$, controlled by $\bar{\alpha}_t$, to obtain $x_t$.
During the reverse process, we directly denoise $x_{t^{\prime}}$ at a sampled $t^{\prime}$ step using the linear transformer to obtain the denoised embeddings $\hat{x_0}$.} 


\pageenlarge{2}
\noindent \textbf{$\bullet$} \textit{Interaction Encoding:}
\looseness -1 As shown in Figure~2, our DiffGT model uses a graph neural recommender to calculate
the initial user/item node representations $x_{0}$ via neighbourhood aggregation according to an adjacency matrix $\mathbf{A}$. 
\xw{To address \textit{\textbf{Limitation 3}}, we argue that incorporating additional data, such as \io{the} side information \zx{of both users and items}, 
could better approximate the interaction graph \io{and become} closer to a `ground truth' graph, \io{thereby addressing the non-informative} \zx{interaction graph}. }
This information includes,
for instance, the \io{genre} of \io{the} movies, 
thereby enriching the matrix with more contextually relevant data.
\xw{Moreover}, our DiffGT model employs a light-weight graph neural recommender~\cite{he2020lightgcn} to aggregate the augmented interaction graph as follows:
\begin{equation}
\mathbf{X}_{G} =\frac{1}{1+\ell_{1}}\left(\mathbf{X}^{0}_{G}+\overline{\mathbf{A}} \mathbf{X}^{0}_{G}+\ldots+\overline{\mathbf{A}}^{\ell_{1}} \mathbf{X}^{0}_{G}\right)
\end{equation}
where $\mathbf{X}_{G}$ denotes the user/item representations after $\ell_{1}$ layers of graph convolution operations, $\overline{\mathbf{A}}$ is the updated adjacency matrix, 
which is enriched by measuring the similarity between user-user and item-item pairs using \io{the} used side information. 
\xw{Hence, introducing $\overline{A}$ \io{aims to enrich} the \io{interaction graph}  while \io{also using} the diffusion process to further reveal \io{the} true user-item interactions.}

\pageenlarge{3}
\noindent \textbf{$\bullet$} \textit{Conditional Denoising Transformer:}
As discussed in Section~\ref{sec:pre}, existing diffusion approaches essentially elaborate an unconditioned denoising paradigm, as outlined as \textit{\textbf{Limitation 2}}.
\zx{To address this limitation, we \io{propose} to refine the reverse process with a condition vector particularly tailored to the recommender system, as shown in Figure~2.}
In our approach, we train a denoising network $f_\theta$ in the reverse diffusion process to denoises the corrupted $\mathrm{x}_t$ and \io{to} \xw{recover} the original interaction vector $\mathrm{x}_0$.
\zx{Then,} we let the denoising model $f_\theta$ directly predict $\mathbf{x}_{0}$ by sampling various $\mathbf{x}_{t}$:
\begin{equation}\label{eqn:condition}
\begin{aligned}
p\left(\mathbf{x}_{0} \mid \mathbf{x}_t, \mathbf{c}\right) & =\mathcal{N}\left(\mathbf{x}_{t} ; {\mu_\theta}_t\left(\mathbf{x}_t, \mathbf{x}_0\right), {\beta}_t \mathbf{I}\right) \\
\end{aligned}
\end{equation}
where $\mathbf{c}$ is the condition, \zx{typically derived from \zx{the} users' interacted items in a \io{recommendation} setting.}
Then this condition is concatenated with \zx{the} noisy user/item \io{embeddings} to predict $\mathbf{x}_{0}$.
\xw{Such personalised interaction information guides the reverse process in an effective manner, which further benefits the downstream recommendation task.}
\xw{Specifically, \io{we emply} a linear transformer~\cite{wang2020linformer}} to denoise the obtained $\mathbf{X}_G$ in the reverse process:
\begin{equation}
\begin{aligned}
\mathbf{X}_T^{\ell_{2}+1} & =\operatorname{LinearAttn}^{\ell_{2}}\left(\mathbf{X}_G^{\ell_{2}},\mathbf{X_{C}}\right), \\
\end{aligned}
\end{equation}
where LinearAttn denotes a linear attention mechanism at the $\ell_{2}$-th layer, $\mathbf{X_{C}}$ is the average embedding of the user's interactions.

\noindent \textbf{$\bullet$} \textit{Efficient Graph Diffusion:}
\update{To explore an effective and efficient diffusion strategy, aside from our diffusion approach, which operates at the embedding level (i.e., continuous diffusion), we conduct a comparative analysis with another variant of a diffusion approach that optimises the adjacency matrix of user-item interactions (i.e., discrete diffusion) \cite{vignac2023digress, haefeli2022diffusion}.}
Figure~3(a) and~3(b) illustrate the \xw{discrete and continuous diffusion methods, respectively.} 
The computation of the posterior distribution in \zx{the} discrete diffusion is as follows:
\begin{equation}\label{eqn:discrete}
\begin{aligned}
    q\left(\mathbf{x}_t \mid \mathbf{x}_{t-1}\right) & = \mathbf{x}_{t-1}\left(\boldsymbol{Q}^{t-1}\right)^{\prime},
    & \text{where } \boldsymbol{Q}^{t}=\left[\begin{array}{cc}1-\alpha_t & \alpha_t\\
\alpha_t & 1-\alpha_t,
\end{array}\right]
\end{aligned}
\end{equation}
where $\boldsymbol{Q}^{\prime}$ is the transpose of $\boldsymbol{Q}$.
$\alpha_t$ is the probability of not changing the edge state\update{s} in the user-item interaction graph.
\zx{Since} the user-item graph is symmetric in recommender system, 
we focus on modelling the upper triangular part of the adjacency matrix.
From Equation~\eqref{eqn:discrete}, the discrete graph diffusion 
involves a series of matrix multiplications throughout the diffusion steps. In contrast, the continuous graph diffusion, as defined in \io{Equation}~\eqref{eqn:forward}, \zx{operates} \zx{an embedding interpolation}, which is deemed to be more efficient than a series of matrix multiplications given a fixed number of diffusion steps. 
In Section~\ref{sec:results}, we present a detailed efficiency analysis on discrete and continuous diffusion processes. 

\pageenlarge{3}
\noindent \textbf{$\bullet$} \textit{Optimisation:}
\looseness -1 \zx{To ensure \io{an} effective denoising from \io{implicit interactions},}
\zx{we optimise our DiffGT model with three loss functions, \io{namely}}
\io{the} Bayesian Personalised Ranking (BPR) \xw{loss} $\mathcal{L}_{BPR}$ for the recommendation task, \xw{the} diffusion loss \xw{$\mathcal{L}_{diff}$} for \xw{optimising} the denoising \xw{outcome} and \io{the} contrastive learning loss $\mathcal{L}_{cl}$: 
\begin{equation}\label{eqn:loss}
    \mathcal{L}  =  \mathcal{L}_{bpr} + \lambda_{1}\mathcal{L}_{diff} + \lambda_{2} \mathcal{L}_{cl},
\end{equation}
The diffusion loss \xw{$\mathcal{L}_{diff}$} is derived from Equation~\eqref{eqn:diffusion}, which regulates the denoising of \io{the} user and item embeddings.
We also aim to maximise the agreement between \io{the} original user/item embeddings and \io{the} denoised user/item embeddings to generate additional supervision signals as follows:
\begin{equation}\label{eqn:cl}
    \mathcal{L}_{cl}=-\log \frac{\exp \left({\mathbf{e}}^{\top} \mathbf{e}^{\prime} / \boldsymbol{\tau}\right)}{\sum_{i=1}^{n} \exp \left({\mathbf{e}}^{\top} \mathbf{r} / \boldsymbol{\tau}\right)}
\end{equation}
where $\mathbf{e}$ is the user/item embedding obtained from a graph encoder, $\mathbf{e}^{\prime}$ is the denoised user/item embedding after $\mathbf{e}_{u}$ \io{has been} processed through the transformer, $\mathbf{r}$ is the embedding of a different user/ item node 
and $\tau$ is a hyper-parameter that adjusts the dynamic range of the resulting loss value.
As such, additional gradients generated from contrastive learning can facilitate effective \io{denoisation} by optimising our graph transformer model in the diffusion process. 
\section{Experiments}
To demonstrate the effectiveness of our DiffGT model, including \xw{its} use of directional noise and the \xw{proposed} graph transformer architecture, we conduct experiments on three real-world datasets to answer the following \io{four} research questions:

\noindent \textbf{RQ1}: \xw{Can our \io{proposed} DiffGT model, leveraging the diffusion process, outperform existing baselines in the top-$k$ recommendation?} 

\noindent \textbf{RQ2}: \xw{What is the performance impact of \io{the various key components of} DiffGT,} 
namely \io{its} directional noise, \io{its} graph transformer, \io{its} augmented adjacency matrix, \io{its} condition and its loss functions?  

\noindent \textbf{RQ3}: Can we efficiently incorporate diffusion \xw{\io{when} making recommendations}?

\xw{We also examine the \io{generalisation} of applying the directional noise diffusion to \io{other} recommendation models, \io{in addition to} the graph neural \io{recommenders}: 
}

\looseness -1 \noindent \textbf{RQ4}: \io{Do our proposed directional diffusion and linear transformer approaches generalise to other recommendation models such as knowledge graph and sequential recommenders?}



\begin{table}[tb]
\caption{Statistics of the used datasets.}
\begin{adjustbox}{width=.9\linewidth}
\setlength{\tabcolsep}{4mm}
\begin{tabular}{c|c | c | c |c c}
\cline{1-5}
\hline
Dataset \ & Users & Items & Interactions & Density\\
\hline
\hline
\textbf{Movielens-1M} &  6,040  &  3,704 & 1,000,209 & 4.47\% \\ 
\hline
\textbf{Foursquare} &  2,060  &  2,876 &  27,149  & 0.46\%  \\ 
\hline
\textbf{Yelp2018} & 31,668  &   38,048 &  1,561,406 & 0.13\% \\ 
\hline
\end{tabular}
\vspace{-4mm}
\end{adjustbox}
\label{tab:dataset}
\end{table}
\noindent \textbf{Datasets:}
\looseness -1 \xw{To answer the research questions RQ1-3 \io{in relation to} the effectiveness and efficiency of DiffGT, we conduct an extensive evaluation on three real-world datasets}, 
namely \textit{Movielens-1M}\footnote{\url{https://grouplens.org/datasets/movielens/1m/}}, \textit{Foursquare}\footnote{\url{https://sites.google.com/site/yangdingqi/home/foursquare-dataset}} and \textit{Yelp2018}\footnote{\url{https://www.yelp.com/dataset}}.
\io{Table~\ref{tab:dataset} presents the} statistics of \io{the three used} datasets.
Following the \xw{experimental setup} of \cite{he2020lightgcn}, we randomly split the aforementioned datasets into training, validation, and testing sets with a 7:1:2 ratio.
\xw{Recall \io{from} the discussion in Section~\ref{sec:diffgt},} \io{that} \xw{we aim to} leverage the abundant side information within these three datasets to augment the user-item interaction graph for effective diffusion. To do so, we follow the methodology of \cite{meng2021graph} \io{for enriching} the adjacency matrix with side information.
\zx{We compute the cosine similarity, which compares two vectors with an inner product, for each pair of user or item feature vectors derived from the side information.} 
\zx{We then identify and select the top-$n$ similar users/items accordingly.} \zx{Next, we update the adjacency matrix by converting \io{a 0} to \io{a 1} for each similar user-user or item-item pair, resulting an enriched outcome.}
\pageenlarge{2}
\noindent \textbf{Baselines:}
\looseness -1 We evaluate the effectiveness of our \io{proposed}  DiffGT\footnote{Code \& scripts will be released upon acceptance of the paper.} model by comparing it with ten \io{existing} state-of-the-art recommendation models. 
These baseline models cover a diverse range of approaches, which include conventional and diffusion\xw{-enhanced} recommenders, side information\xw{-enriched} recommenders and \zx{graph neural recommenders}:
\xw{\textbf{(1) \textit{Conventional approaches}}}: \textbf{MF~\cite{tang2016cross}} decomposes the user-item interaction matrix into the product of two low-rank matrices, representing the latent factors of \io{the} users and \io{items,} respectively;
\noindent \textbf{CDAE~\cite{wu2016collaborative}} introduces random noises to \io{the} user-item interactions and uses an auto-encoder for denoising the implicit interactions;
\noindent \textbf{MultiVAE~\cite{liang2018variational}} \io{uses} Variational Auto-Encoders (VAEs) with multinomial likelihood to model implicit interactions;
\xw{\textbf{(2) \textit{Diffusion-enhanced recommender}}}:
\noindent \textbf{DiffRec~\cite{wang2023diffusion}} is a VAE-based diffusion approach 
that infers \io{the} users’ preferences by modelling the interaction probabilities in a denoising manner;
\xw{\textbf{(3) \textit{Side information-enriched models}}}:
\noindent \textbf{HIRE~\cite{liu2019recommender}} is a side information-enhanced recommender, which combines the flat and hierarchical side information with implicit interactions to enhance the recommendation performance;
\noindent \textbf{cVAE~\cite{chen2018collective}} encodes the side information and the user ratings via VAE for an enhanced performance;
\xw{\textbf{(4) \textit{Graph neural recommenders:}}}
\noindent \textbf{LightGCN~\cite{he2020lightgcn}} is a light \zx{graph neural recommender}, characterised by the removal of non-linear activation functions and \io{the} transformation matrices in the feature propagation process.
This approach has been widely used as a strong graph recommender \io{for} modelling implicit interactions for top-$k$ recommendation~\cite{gao2022graph};
\noindent \textbf{SGL~\cite{wu2021self}} leverages contrastive learning to enhance LightGCN. It conducts different augmentation operators such as edge dropout and node dropout to generate multiple samples for subsequent contrasting;
\noindent \textbf{SimGCL~\cite{yu2022graph}}
uses a noise-based augmentation technique on the user/item embedding level to model the \io{users' preferences} from implicit interactions, resulting in \io{an} improved recommendation performance \zx{when} based on LightGCN;
\noindent \textbf{GFormer~\cite{li2023graph}} is a recently proposed graph transformer model to model implicit interactions, which contrasts the embeddings of a graph encoder and a separate transformer module by selectively dropping out the low-weight edges. 


\noindent \textbf{Experimental Settings:}
All experiments are conducted on a machine equipped with an RTX A6000 GPU for a fair comparison. 
For the hyper-parameters specific to the loss function of DiffGT, \zx{as introduced in Equation~\eqref{eqn:loss}}, we tune each of $\lambda_{1}$ \io{and} $\lambda_{2}$ within the ranges of $\left \{ 0, 0.1,0.2,...,1.0 \right \}$. We also tune $\tau$, \zx{as described in Equation~\eqref{eqn:cl}}, within the ranges of $\left \{ 0, 0.1,0.2,...,1.0,2.0,...10.0 \right \}$. 
For \io{the} other hyper-parameters \zx{ of DiffGT}, \io{which are} optimised on the validation set, 
we tune \zx{the directional noise factor} $\epsilon$ in Equation~\eqref{eqn:noise3}
within the ranges of ~$\left \{ 0, 0.1,0.2,...,1.0, 2.0,...10.0 \right \}$, \io{the} diffusion step $t$ \zx{within} the range of $\left \{ 10, 20, ..., 100 \right \}$, 
\zx{the number of top-$n$ similar neighbours for a user or item within the range of $\left \{ 0, 5, ..., 20 \right \}$}
and the \zx{number of layers in the} transformer layer \zx{varies over} $\left \{ 1,2, ...,5 \right \}$.
For the used baselines, 
\zx{we employ a grid search method to tune all the hyper-parameters. \io{We ensure} that the ranges of parameters or the settings, such as \io{the} learning rate ($\left \{10^{-2},10^{-3},10^{-4}  \right \}$), \io{the} batch size (2048), and \io{the} initialisation (Xavier) are consistent with the ones used in our DiffGT model, \io{so as to enforce} a fair comparison.}
In contrast to previous studies that rely on a single oracle testing set per dataset, we construct 10 different testing sets for each dataset using different random seeds.
\zx{This multiple testing setups prevent any single test from favouring specific model characteristics, thus \io{reducing} the evaluation bias~\cite{qian2021my}.}
Hence, the reported performance of each run represents the average of the 10 testing sets, \update{further enforcing a fair comparison.}
Moreover, we adopt two widely used evaluation metrics, namely Recall@$k$ and NDCG@$k$ to evaluate the performance of top-$k$ \io{recommendation} with the \io{$k$} cutoff set to 20.
We also apply early-stopping \zx{for both our DiffGT model and the used baselines}, which terminates the training when the validation loss fails to decrease for 50 epochs.

\section{RESULTS AND ANALYSIS}\label{sec:results}
\pageenlarge{2}
\begin{table}[tb]
\centering
\caption{Experimental results between our DiffGT model and the used baselines. The best performance of each model is highlighted in bold. $^{*}$ denotes a significant difference compared to the result of baselines using the Holm-Bonferroni corrected paired t-test with $p<0.05$. }
\label{tab:comp_base}
\begin{adjustbox}{width=\linewidth}
\begin{tabular}{lcccccccc}
\toprule
\multirow{1}{*}{\textbf{Dataset}} & \multicolumn{2}{c}{Movielens-1M} & \multicolumn{2}{c}{Foursquare} & \multicolumn{2}{c}{Yelp2018}\\ 
\cmidrule(lr){1-1} \cmidrule(lr){2-3} \cmidrule(lr){4-5} \cmidrule(lr){6-7}
Methods & Recall@20 & NDCG@20 & Recall@20 & NDCG@20  & Recall@20 & NDCG@20\\
\midrule
MF & ${0.1638}^{*}$ & ${0.2074}^{*}$ &  ${0.2430}^{*}$ & ${0.3424}^{*}$  &  ${0.0409}^{*}$ & ${0.0341}^{*}$ \\
CDAE & ${0.2075}^{*}$ & ${0.2346}^{*}$ &  ${0.2915}^{*}$ & ${0.3877}^{*}$  & ${0.0435}^{*}$ & ${0.0363}^{*}$\\
MultiVAE & ${0.2740}^{*}$  & ${0.3085}^{*}$  &  ${0.4151}^{*}$ & ${0.5580}^{*}$ & ${0.0557}^{*}$ & ${0.0481}^{*}$\\
\midrule
DiffRec & ${0.2756}^{*}$  & ${0.3108}^{*}$  &  ${0.4414}^{*}$ & ${0.6087}^{*}$ & ${0.0663}^{*}$ & ${0.0551}^{*}$\\
\midrule
cVAE & ${0.2243}^{*}$  & ${0.2665}^{*}$  &  ${0.3604}^{*}$ & ${0.4747}^{*}$ & ${0.0483}^{*}$ & ${0.0417}^{*}$\\
HIRE & ${0.2709}^{*}$  & ${0.3050}^{*}$  &  ${0.3634}^{*}$ & ${0.4891}^{*}$ & ${0.0613}^{*}$ &  ${0.0520}^{*}$ \\
\midrule
LightGCN & {{0.2697}$^{*}$} & ${0.3021}^{*}$ &  ${0.4161}^{*}$ & ${0.5707}^{*}$  &  $0.0589^{*}$ &  $0.0504^{*}$\\
SGL & ${0.2723}^{*}$  & ${0.3078}^{*}$ & {0.4333}$^{*}$ & {0.5978}$^{*}$ & ${0.0646}^{*}$ & ${0.0537}^{*}$ \\
SimGCL & ${0.2788}^{*}$  & ${0.3112}^{*}$ & {0.4353}$^{*}$ & {0.6214}$^{*}$ & ${0.0701}^{*}$ & ${0.0573}^{*}$ \\
GFormer & ${0.2812}^{*}$  & ${0.3142}^{*}$ & {0.4242}$^{*}$ & {0.6051}$^{*}$ & ${0.0696}^{*}$ & ${0.0559}^{*}$ \\
\midrule
DiffGT & $\textbf{0.2903}$  & $\textbf{0.3264}$ & \textbf{0.4589} & \textbf{0.6612} & $\textbf{0.0715}$ & $\textbf{0.0587}$ \\
\bottomrule
\end{tabular}
\vspace{-4mm}
\end{adjustbox} 
\end{table}

\noindent \textbf{Overall Performance (RQ1):}
Table~\ref{tab:comp_base} reports the results of \io{comparing} our DiffGT model  \io{with} all the \io{used} baselines. \\
\noindent \textbf{$\bullet$} \textit{Performance of DiffGT:}
\looseness -1 \xw{\io{From the observed} results on three} datasets, \io{we can conclude that} DiffGT significantly outperforms the baselines \io{on both uses metrics} \io{as confirmed by a} paired t-test with \io{the} Holm-Bonferroni correction. 
The improvements can be attributed to the integration of a graph transformer network that incorporates the directional noise within the forward process, \xw{and its denoising process, which is conditioned on personalised information \zx{from the user's historical interactions}.}
\\
\noindent \textbf{$\bullet$} \textit{Effectiveness of \io{the} Transformer Architecture:}
We also evaluate
the graph transformer network in DiffGT by comparing its performance to that of classical graph neural recommenders (LightGCN, SGL, SimGCL) and another graph transformer \io{method} (GFormer). 
Table~\ref{tab:comp_base} shows that DiffGT \xw{and} GFormer, \xw{both \io{include a}  graph transformer} architecture, outperform the classical \zx{graph neural recommenders} (LightGCN, SGL, SimGCL). 
\xw{These results show} the effectiveness of the graph transformer network in 
\xw{modelling \io{the users'} preferences}.
\io{Note that} GFormer uses the transformer as a parallel encoder to construct a view for both \io{the} user and item \io{embeddings} in a contrastive loss. \xw{In contrast,} our DiffGT model leverages a cascading architecture of \zx{a} graph encoder and \zx{a} transformer to effectively facilitate the denoisation in the diffusion process. The \io{difference in performance between DiffGT and} GFormer illustrates the superiority of our cascading architecture.\\
\noindent \textbf{$\bullet$} \textit{Effectiveness of the Diffusion-enhanced Recommenders:} 
\zx{We \io{now} discuss the effectiveness of the diffusion recommenders (DiffRec, DiffGT) by comparing their performance to the conventional models (MF, CDAE, MultiVAE).}
\zx{\io{From} Table~\ref{tab:comp_base}, we observe that the diffusion recommenders (DiffGT and DiffRec) significantly outperform the conventional recommenders (MF, CDAE, MultiVAE).}
\zx{The \io{superiority of the} diffusion recommenders \io{confirms} the effectiveness of the diffusion process in effectively denoising the implicit interactions.}
\zx{Moreover, Table~\ref{tab:comp_base} shows that our DiffGT model significantly outperforms DiffRec, which demonstrates the effectiveness of our unique diffusion process in injecting a directional noise via a graph transformer architecture rather than using a normal Gaussian noise.}
\noindent \textbf{$\bullet$} \textit{Effectiveness of Leveraging More Noisy Interactions:}
\io{By comparing DiffGT with} the methods \io{that} also leverage side information, \io{we observe from} Table~\ref{tab:comp_base} that DiffGT significantly outperforms the side information-enhanced methods (cVAE, HIRE) by a large margin. 
This result indicates that DiffGT is \io{indeed \zx{more} effective} at \update{using} the
\zx{\zx{side} information, such as \update{movie genres}}, than cVAE and HIRE.

\looseness -1 In answer to \textbf{RQ1}, we conclude that our DiffGT model successfully leverages the directional noise in a graph transformer architecture for an effective diffusion process and outperforms all the existing strong baselines.

\label{ss:ablation}
\begin{table}[tb]
\centering
\caption{Ablation study on key components of DiffGT. $^{*}$ denotes a significant difference compared to the result of baselines using paired t-test with $p<0.05$.}
\label{tab:ablation}
\begin{adjustbox}{width=\linewidth}
\begin{tabular}{lcccccccc}
\toprule
\multirow{1}{*}{\textbf{Dataset}} & \multicolumn{2}{c}{Movielens-1M} & \multicolumn{2}{c}{Foursquare} & \multicolumn{2}{c}{Yelp2018}\\ 
\cmidrule(lr){1-1} \cmidrule(lr){2-3} \cmidrule(lr){4-5} \cmidrule(lr){6-7}
Variants & Recall@20 & NDCG@20  & Recall@20 & NDCG@20  & Recall@20 & NDCG@20\\
\midrule
-Direction & ${0.2533}^{*}$ & ${0.2843}^{*}$ &  ${0.4016}^{*}$ & ${0.5029}^{*}$  &  $0.0541^{*}$ & ${0.0474}^{*}$\\
-Condition & ${0.2822}^{*}$ & ${0.3145}^{*}$ &  ${0.4322}^{*}$ & ${0.5978}^{*}$  &  $0.0703^{*}$ & ${0.0578}^{*}$\\
-Transformer & ${0.2535}^{*}$ & ${0.2967}^{*}$ &  ${0.3631}^{*}$ & ${0.5029}^{*}$  &  $0.0616^{*}$ & ${0.0507}^{*}$\\
-Side & ${0.2858}$ & ${0.3203}$ &  ${0.4352}^{*}$ & ${0.6154}^{*}$  &  $0.0704^{*}$ & ${0.0556}^{*}$\\
\midrule
-CL & ${0.2701}^{*}$  & ${0.3007}^{*}$  &  ${0.4144}^{*}$ & ${0.5609}^{*}$ & ${0.0603}^{*}$ &  ${0.0508}^{*}$ \\
-DiffL & ${0.2736}^{*}$ & ${0.3084}^{*}$ &  ${0.4263}^{*}$ & ${0.5803}^{*}$  &  $0.0616^{*}$ & ${0.0514}^{*}$\\
\midrule
DiffGT & $\textbf{0.2903}$  & $\textbf{0.3264}$ & \textbf{0.4589} & \textbf{0.6612} & $\textbf{0.0715}$ & $\textbf{0.0587}$ \\
\bottomrule
\end{tabular}
\vspace{-10mm}
\end{adjustbox}
\end{table}

\vspace{+2mm}
\pageenlarge{3}
\noindent \textbf{Ablation Analysis (RQ2):}
In this section, we ablate each of the key components of DiffGT, namely \io{the} directional noise \zx{(discussed in Section~\ref{sec:directional})} and the diffusion architecture components \zx{(Section~\ref{sec:diffgt})}, including the conditioned denoising, enriched user-item interaction graph, graph-adapted diffusion architecture and the loss functions. \\
\looseness -1 \noindent \textbf{$\bullet$} \textit{Directional Noise:}
As \io{mentioned} in Section~\ref{sec:intro} and highlighted under \textbf{\textit{Limitation 1}} \xw{(see Section~\ref{sec:pre})}, recommendation \io{data} often exhibits unique anisotropic and directional structures. 
\update{To enable a gradual addition of noise, and the recovery of the original embeddings in the denoising procedure,}
we \io{use a} directional Gaussian noise in the forward process \zx{of DiffGT}.
To demonstrate the effectiveness of \io{this} directional noise, we \io{conduct} a comparative analysis by replacing \io{the} directional Gaussian noise with \io{a} normal Gaussian noise. This substitution \io{corresponds to the row indicated by ``-Direction''} in Table~\ref{tab:ablation}.
From Table~\ref{tab:ablation}, we observe that DiffGT significantly \io{outperforms} the ``-Direction'' variant. This \xw{finding} further confirms the effectiveness of incorporating directional noise, \io{so as to align} with the inherent directional structure of the recommendation data, thereby enhancing the recommendation performance.

\begin{figure}[tb]
    \begin{subfigure}[t]{0.49\linewidth}
        \includegraphics[trim={3.8cm 3.5cm 3.8cm 3.5cm},clip,width=1\linewidth]{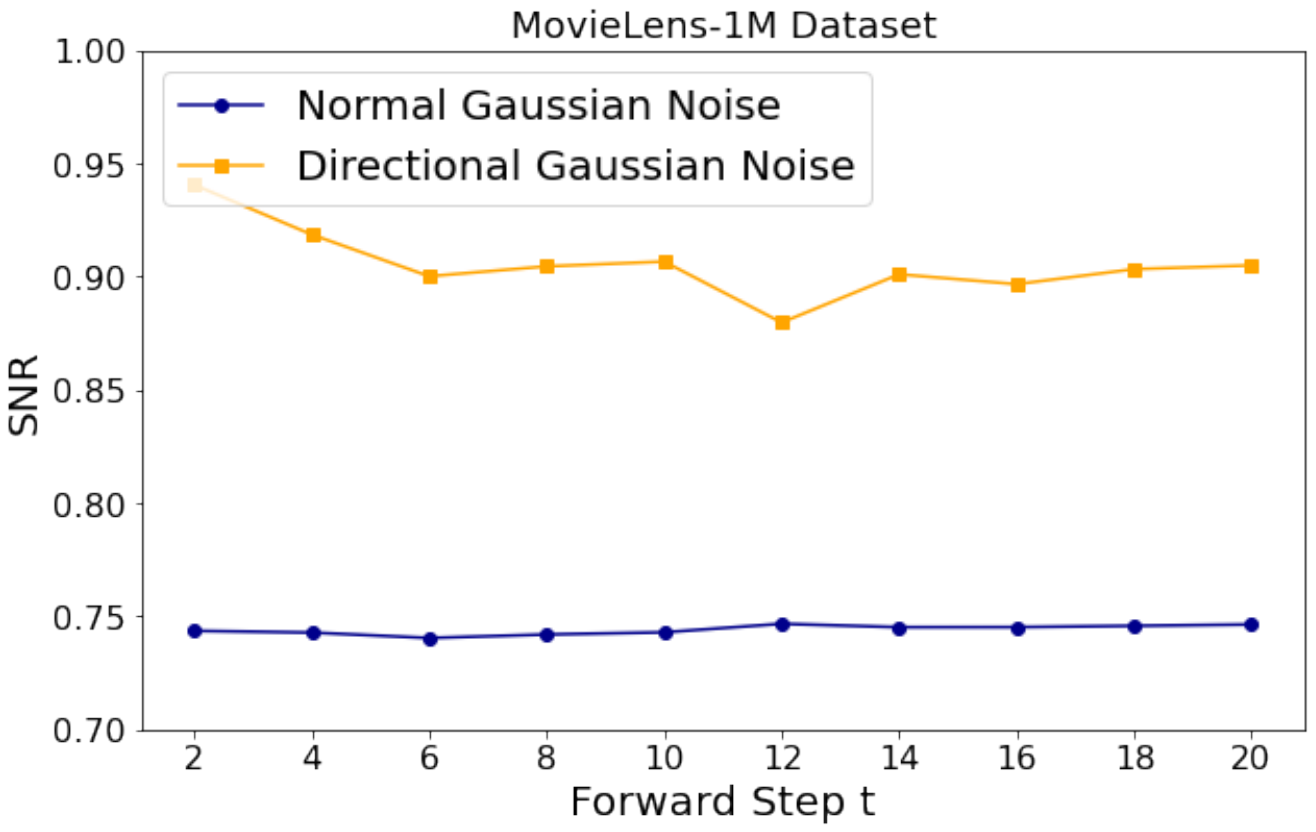}
        \captionsetup{labelformat=empty}
    \end{subfigure}
    \begin{subfigure}[t]{0.49\linewidth}
        \includegraphics[trim={3.8cm 3.5cm 3.8cm 3.5cm},clip,width=1\linewidth]{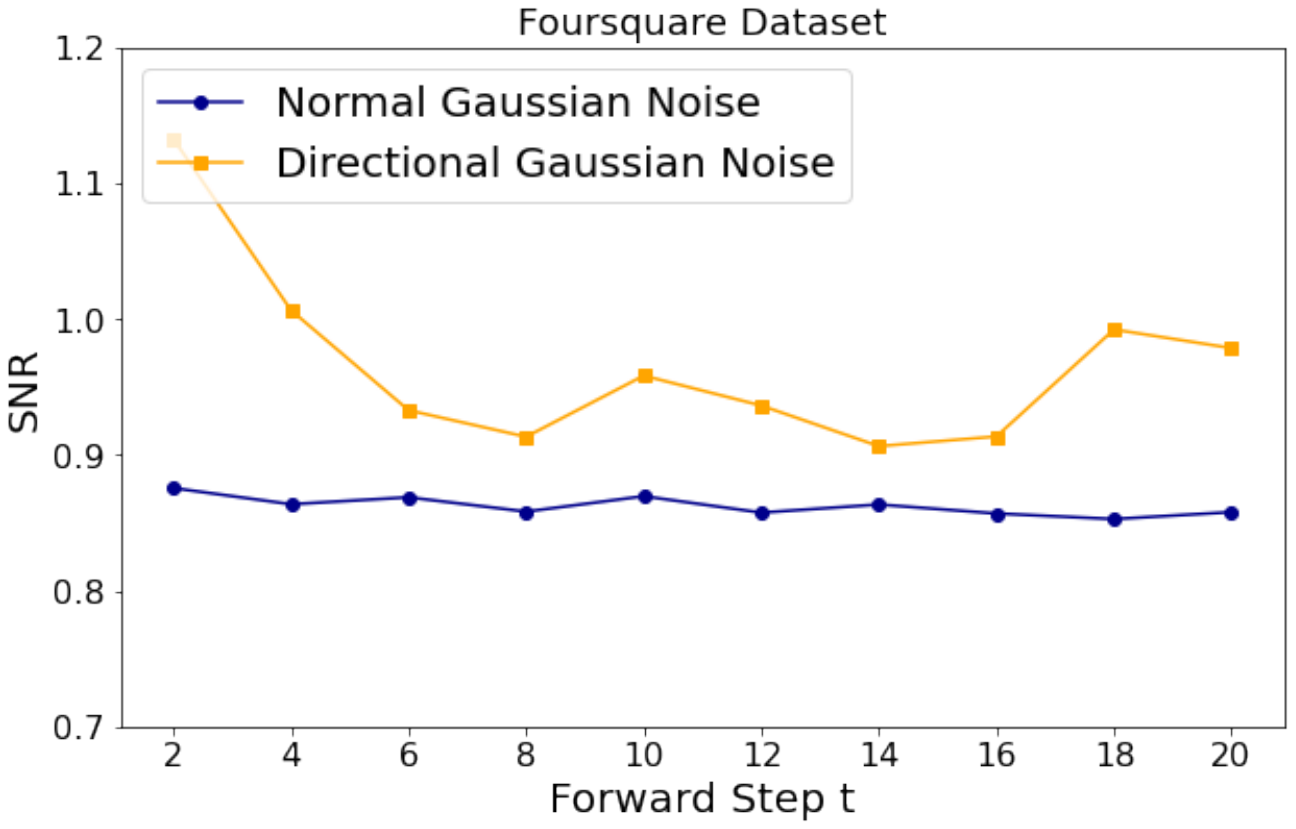}
        \captionsetup{labelformat=empty}
    \end{subfigure}
    \captionsetup{labelformat=simple}
    \vspace{-2mm}
    \caption{The Signal-to-Noise Ratio (SNR) curves along different diffusion steps on MovieLens-1M and Foursquare.}
\label{fig:snr_noise}
\vspace{-4mm}
\end{figure}

\pageenlarge{3}

\looseness -1 To further explore the impact of incorporating \io{a} normal Gaussian noise \io{or a} directional Gaussian noise in our DiffGT model, 
we calculate the Signal-to-Noise Ratio (SNR) at each forward step of the diffusion process using Linear Discriminant Analysis (LDA)~\cite{xanthopoulos2013linear}. 
Specifically, we calculate the SNR \io{values} by encoding the noisy user/item embeddings with a pre-trained \zx{graph neural recommender}~\cite{he2020lightgcn} and project these embeddings into a linearly separable space.
These SNR values quantify the informativeness of the learned representations at different steps of the diffusion process.
Figure~\ref{fig:snr_noise} shows notable \io{differences} in SNR values between using \io{the} normal Gaussian noise and \io{the} directional Gaussian noise on the MovieLens-1M and Foursquare \io{datasets}.
We omit the results of the Yelp2018 dataset \xw{due to the page limit \io{but note that we} observe similar \io{trends and conclusions}} across all three datasets.
\io{Figure}~\ref{fig:snr_noise} \io{particularly shows} that using \io{a} normal Gaussian noise in our DiffGT model consistently results in \zx{lower} SNR values compared to \io{the usage of a} directional Gaussian noise.
\io{The notable low SNR values indicate} that there is a high noise level within the embeddings, thus leading to a denoisation difficulty in the subsequent reverse process.
\zx{In contrast, the SNR values of using a directional noise exhibit a slight decrease while still maintaining \io{higher values} compared to the normal Gaussian noise,}
\zx{which means 
the usage of a directional noise brings varied noise levels within the embeddings during the diffusion process.}
\update{The variation in noise levels enables our DiffGT model to adaptively} \zx{learn to denoise from different noise strengths of embeddings, thus enabling a more effective diffusion process.}
\zx{\update{This} observed difference of SNR values also aligns with Equation~\eqref{eqn:noise2}, which constraints the added noise to share the same coordinate sign with the embeddings $x_{0}$, thereby indicating its data-dependant characteristic.}
\zx{Consequently, our DiffGT model effectively enables the reverse process by looking into the noisy embeddings in different steps while considering the preserved information at different levels.}\\
\noindent \textbf{$\bullet$} \textit{Conditioned Denoising:}
As discussed in Section~\ref{sec:pre} \io{and} \textbf{\textit{Limitation 2}}, existing diffusion methods in recommender systems essentially \zx{adopt} an uncondition\zx{ed} generation paradigm.
To address this limitation, our DiffGT model \zx{uses} the average embedding of \io{the} user interactions as a condition to guide the diffusion process.
To examine the effectiveness of conditioning \io{the} diffusion \zx{over an unconditioned diffusion process},
we remove the condition $C$ \zx{in Equation~\eqref{eqn:condition}} during the reverse process \io{of DiffGT} and compare its recommendation performance with our original DiffGT model.
We observe from Table~\ref{tab:ablation} that the ``-Condition'' variant underperforms DiffGT, which includes a condition \zx{using the average embedding of the user interactions}.
This result indicates that the average embedding of \io{the} user interactions effectively guides the reverse process to \zx{enable a more effective diffusion process.}\\
\noindent \textbf{$\bullet$} \textit{Enhanced User-item Interaction Graph:}
As discussed in Section~\ref{sec:diffgt} \io{and} \textbf{\textit{Limitation 3}}, we aim to incorporate additional data, such as \io{the} side information, to \zx{bring} the interaction graph \zx{closer} to a ‘ground truth’ graph.
To examine the \io{impact} of incorporating side information \io{on the effectiveness of} diffusion, 
we conduct a comparative analysis of our DiffGT model's performance with and without an adjacency matrix enhanced by side information.
In Table~\ref{tab:ablation}, the variant without side information is denoted as ''-Side''. 
The results \io{show} that our side information-enhanced DiffGT model significantly outperforms the "-Side" variant in four out of six instances.
\zx{Overall, the results show that the use of side information is a promising approach for further enhancing the denoising diffusion process in top-$k$ recommendation.}\\
\pageenlarge{3}
\noindent \textbf{$\bullet$} \textit{\update{Graph-adapted Diffusion Architecture:}}
\textbf{\textit{Limitation 4}} in Section~\ref{sec:diffgt} \zx{stipulates} that the typical architectures used in \io{the} diffusion models are not compatible with \io{the} \zx{graph neural recommenders}.
To examine the effectiveness of our \io{proposed} graph transformer architecture \zx{in denoising the implicit interactions}, we substitute the transformer with a weighted matrix to perform denoising since the weighted matrix is the commonly used method for denoising \io{in a graph VAE architecture}~\cite{kong2023autoregressive}.
\xw{Table~\ref{tab:ablation}} \io{shows} that our DiffGT model outperforms the ``-Transformer'' variant with the weighted matrix by a large margin. This superior performance demonstrates the \io{suitability} of a transformer network \zx{in enhancing the} diffusion model for \zx{top-$k$} recommender systems. \\
\noindent \textbf{$\bullet$} \textit{Loss Functions:}
As shown in Table~\ref{tab:ablation}, 
the inclusion of both \io{the} diffusion loss ( the ``-DiffL'' variant) and \io{the} contrastive loss (the ``-CL'' variant) significantly enhances the performance of our DiffGT model.
We attribute these improvements to the effect of the diffusion loss \zx{and} the \zx{added-value of the} contrastive loss in effectively denoising the user/item embeddings from \io{the} implicit interactions.

\io{Overall,} in response to RQ2, \io{we conclude that} our DiffGT model successfully leverages each of its components as well as its loss functions. 
\zx{Notably, the incorporation of the directional noise and the linear transformer significantly contribute to enhancing the model's performance in the top-$k$ recommendation task.}
We \io{have also} conducted a hyper-parameter sensitivity analysis on our DiffGT model, 
demonstrating that its performance remains relatively stable across the variations of these hyper-parameters. However, due to page constraints, these details are not included \io{in this paper}.

\begin{table}[tb]
\centering
\vspace{+1mm}
\caption{\io{Comparative} time complexity. $^*$ and $^+$ indicate paired t-test and TOST test results \zx{using the Holm-Bonferroni correction} with $p < 0.05$, respectively.}
\label{tab:efficiency}
\begin{adjustbox}{width=\linewidth}
\begin{tabular}{lcccc}
\toprule
\multirow{1}{*}{\textbf{Diffusion Process}} & \multicolumn{1}{c}{Forward Process} & \multicolumn{1}{c}{Reverse Process} & \multicolumn{1}{c}{Recall@20} & \multicolumn{1}{c}{NDCG@20}\\ 
\midrule
DiffGT (discrete) & $\mathcal{O}({T}\cdot{2E}\cdot{M^2}\cdot{L_{1}d})$ & $\mathcal{O}({T}\cdot{N^2}\cdot{L_{2}d})$ &  $0.2657^{*}$ & ${0.2873}^{*}$\\
DiffGT (continuous) & $\mathcal{O}({T}\cdot{2E}\cdot{L_{1}d})$ & $\mathcal{O}({T}\cdot{N^2}\cdot{L_{2}d})$ &  $0.2917^{}$ & ${0.3269}$\\
\midrule
DiffGT (continuous-linear) & $\mathcal{O}({T}\cdot{2E}\cdot{L_{1}d})$ & $\mathcal{O}({T}\cdot{N}\cdot{L_{2}d})$ &   $0.2895^{+}$ & ${0.3253}^{}$\\
DiffGT (continuous-sampling) & $\mathcal{O}({T}\cdot{2E}\cdot{L_{1}d})$ & $\mathcal{O}(K\cdot{N^2}\cdot{L_{2}d})$ &   $0.2910^{+}$ & ${0.3270}^{+}$\\
DiffGT & $\mathcal{O}({T}\cdot{2E}\cdot{L_{1}d})$ & $\mathcal{O}({N}\cdot{L_{2}d})$ &   $0.2903^{+}$ & ${0.3264}^{+}$\\
\bottomrule
\end{tabular}
\end{adjustbox}
\vspace{-4mm}
\end{table}

\vspace{+2mm}
\noindent \textbf{Model Efficiency (RQ3):}
In this section, we analyse the time complexity of our DiffGT model, and compare it 
against its variants 
using a discrete diffusion, reverse sampling and the linear transformer, \update{as per the batch time complexity.}
Let $T$ \io{denotes} the \zx{number of steps} in the diffusion process, $E$ is the \zx{number of edges} in the user-item bipartite graph, \io{and} $N$ is the number of user and item nodes. For the forward process, $L_{1}$ is the \zx{number of layers in} the graph encoder and $M$ is the size of a \zx{the} transition matrix for \io{a} discrete diffusion. \io{Since} we leverage a transformer network for \io{an} effective denoising in the reverse process, $L_{2}$ and $d$ represent the \zx{number of layers} and \io{the} feature dimensionality in the used transformer, respectively.
$K$ denotes the number of the sampled steps, where $K << T$.\\
\noindent \textbf{$\bullet$} \textit{Discrete vs. continuous diffusion:}
\zx{As discussed in Section~\ref{sec:intro}, there are two primary approaches in diffusion models: a discrete diffusion, which is tailored for discrete data such as adjacency matrices, and a continuous diffusion which is suitable for the embeddings.}
\zx{Besides applying a continuous diffusion at the embedding level, it is necessary to examine and compare the effectiveness and efficiency of a discrete diffusion to identify the most appropriate approach within the context of recommender systems.}
\zx{In this section, we compare the time complexity and the recommendation performance between using a discrete diffusion (as detailed in Equation~\eqref{eqn:discrete}) and a continuous diffusion (as detailed in Equation~\eqref{eqn:forward})}.
Table~\ref{tab:efficiency} \io{presents} the comparison of the time complexity and recommendation performance
between \zx{the use of a discrete and a continuous diffusion in our DiffGT model, respectively.}
Comparing \io{the} discrete and continuous diffusion, a transition matrix is not required for continuous diffusion. 
As a result, \io{the} continuous diffusion is $M^2$ faster than \io{the} discrete diffusion in the forward process.
\zx{\io{In addition}, \update{Table~\ref{tab:efficiency} shows that} the continuous diffusion outperforms the discrete diffusion in terms of effectiveness on both used metrics. For instance, its recall value of 0.2917 is higher than the 0.2657 of the discrete diffusion.}
\zx{This result shows the continuous diffusion \io{to be} the \io{better suited} approach for DiffGT, as it enhances the recommendation performance while also reducing time complexity.}\\
\pageenlarge{3}
\looseness -1 \noindent \textbf{$\bullet$} \textit{Efficient denoising:}
\zx{Recall from the conventional reverse process in the out-of-box diffusion baseline (i.e., DiffRec) in Equation~\eqref{eqn:reverse},}
\zx{\zx{that} we aim to verify the efficiency and effectiveness of our proposed sampled denoising method as well as the used linear transformer in DiffGT.}
\zx{We examine if these components, outlined in Section~\ref{sec:diffgt}, indeed enable a more efficient reverse process.}
\zx{Table~\ref{tab:efficiency} presents various DiffGT variants:}
\update{DiffGT (continuous) incorporating a vanilla transformer across all diffusion steps},
\zx{DiffGT (continuous-linear) employing a linear transformer, DiffGT (continuous-sampling) utilising a sampled denoising method \update{with reduced steps}, and a combination of both in the \update{original} DiffGT.}
We employ a two one-sided equivalence test (TOST)~\cite{schuirmann1987comparison} on the above variants to ascertain an effectiveness equivalence with the DiffGT (continuous) 
variant.
${\space}$\\
\zx{Table~\ref{tab:efficiency} shows that the DiffGT~(continuous-sampling) variant, compared to the DiffGT (continuous) variant, reduces the time complexity in the reverse process by $T - K$ times.}
\zx{Despite this reduction, the performance of DiffGT (continuous-sampling)} remains on \io{a} par with the DiffGT (continuous) variant.
\zx{This finding suggests that the sampled denoising method maintains an effectiveness equivalent to the conventional approach while being more efficient.}

\zx{We also compare the DiffGT~(continuous-linear) with the DiffGT~(continuous) variant in Table~\ref{tab:efficiency}.}
\zx{We can observe from Table~\ref{tab:efficiency} that the DiffGT~(continuous-linear) variant reduces the quadratic complexity of the number of user/item nodes $N^2$ in the DiffGT~(continuous) variant to a linear factor $N$ during the reverse process}
\zx{while maintaining a competitive effectiveness with the DiffGT~(continuous) variant.}
\zx{This result demonstrates that our DiffGT model successfully employs a linear transformer to enable an efficient and effective reverse process.}
\zx{Similarly, we can obtain the same conclusion when examining the combined use of the sampled denoising method and the linear transformer, as reflected in Table~\ref{tab:efficiency}.}

\zx{Hence, in answer to RQ3, we conclude that the continuous diffusion is the most appropriate diffusion method for the recommendation data. Moreover, our DiffGT model successfully enables an efficient and effective reverse process with our proposed sampled denoising method and the linear transformer.}

\begin{figure}[tb]
    \begin{subfigure}[t]{0.49\linewidth}
        \includegraphics[trim={1.5cm 2.2cm 1cm 2cm},clip,width=1\linewidth]{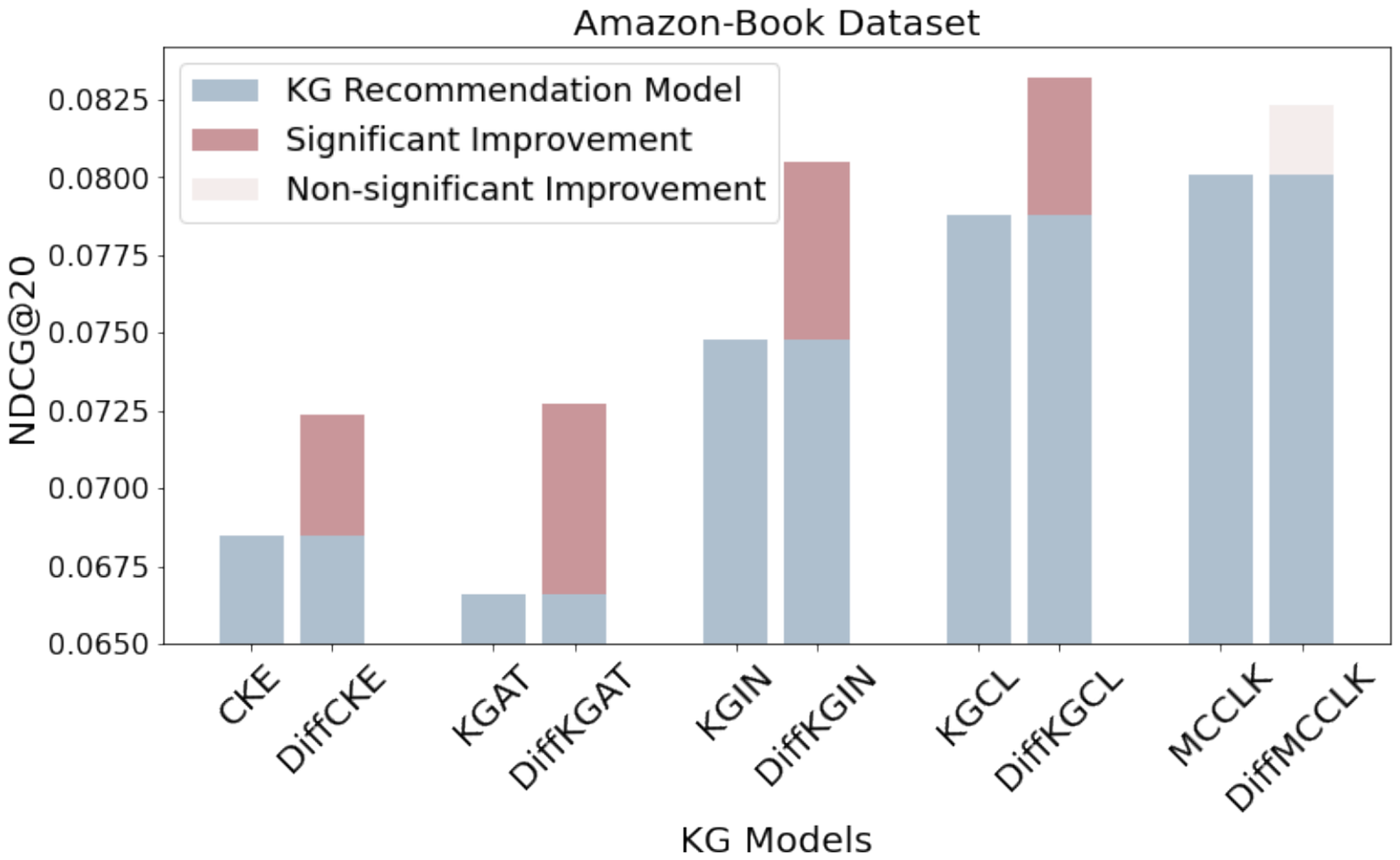}
    \end{subfigure}
    \begin{subfigure}[t]{0.49\linewidth}
        \includegraphics[trim={1.5cm 2.2cm 1cm 2cm},clip,width=1\linewidth]{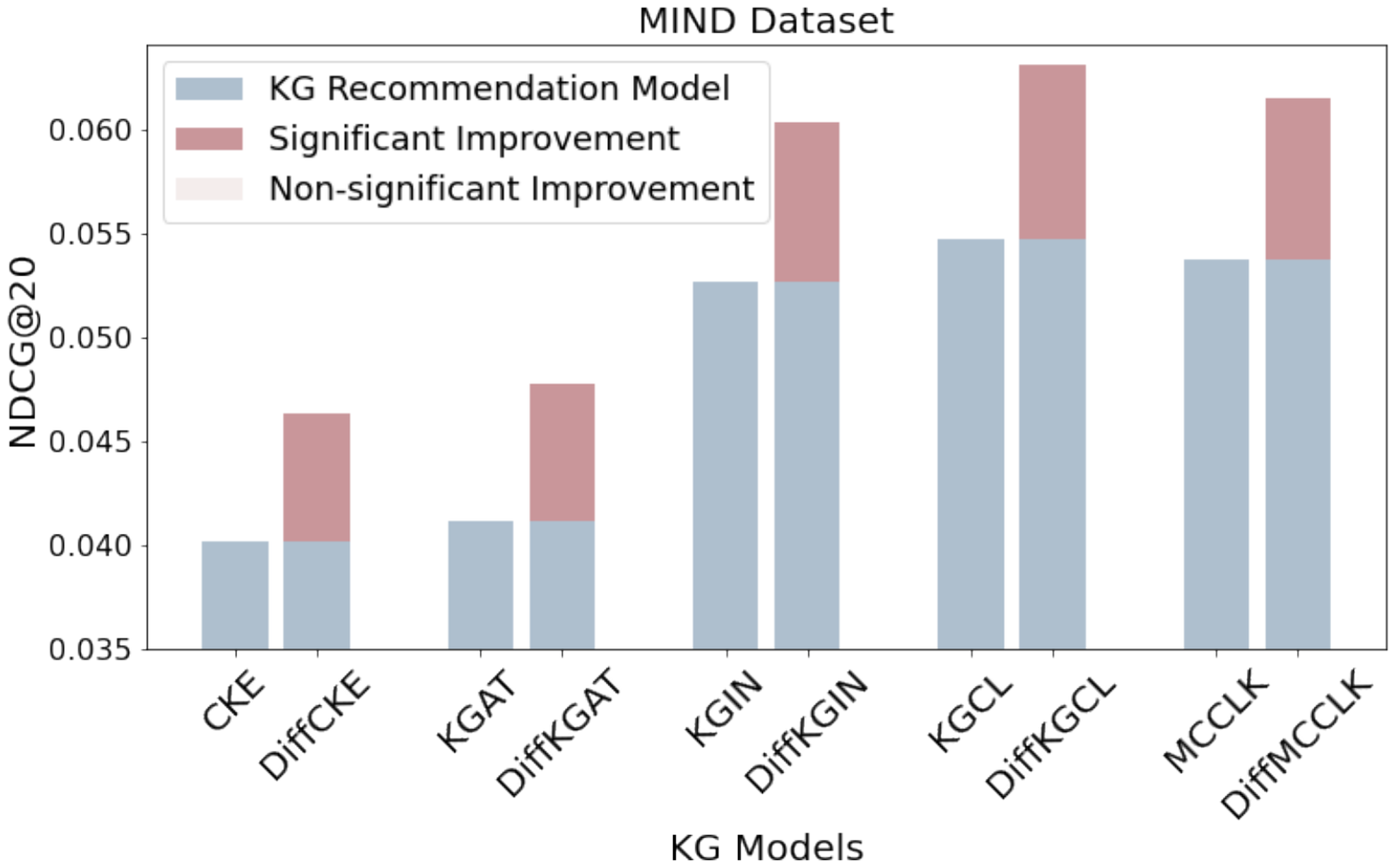}
    \end{subfigure}
    \vspace{+2mm}
    \begin{subfigure}[t]{0.49\linewidth}
        \includegraphics[trim={1.5cm 2cm 1cm 2cm},clip,width=1\linewidth]{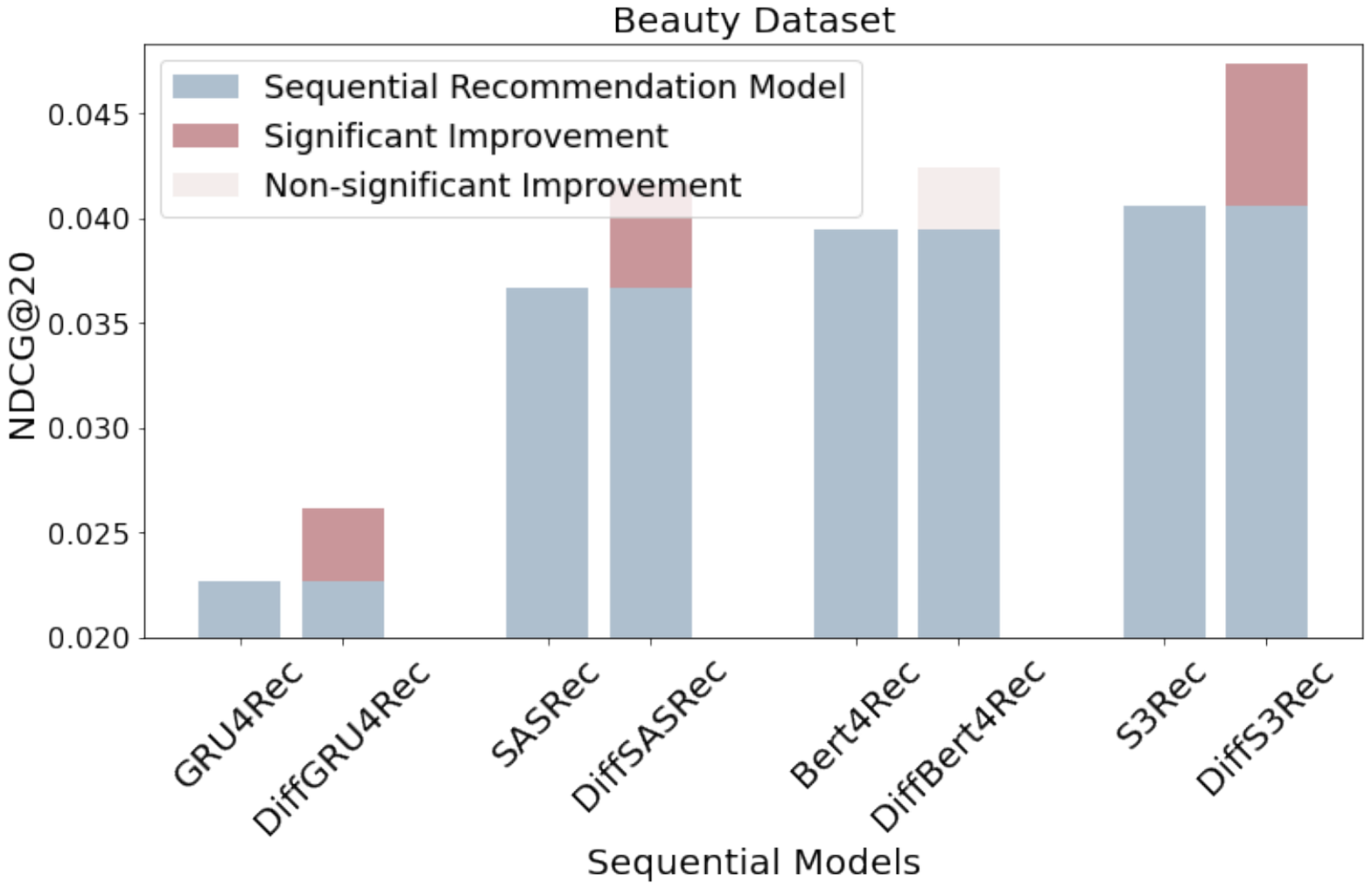}
    \end{subfigure}
    \begin{subfigure}[t]{0.49\linewidth}
        \includegraphics[trim={1.5cm 2cm 1cm 2cm},clip,width=1\linewidth]{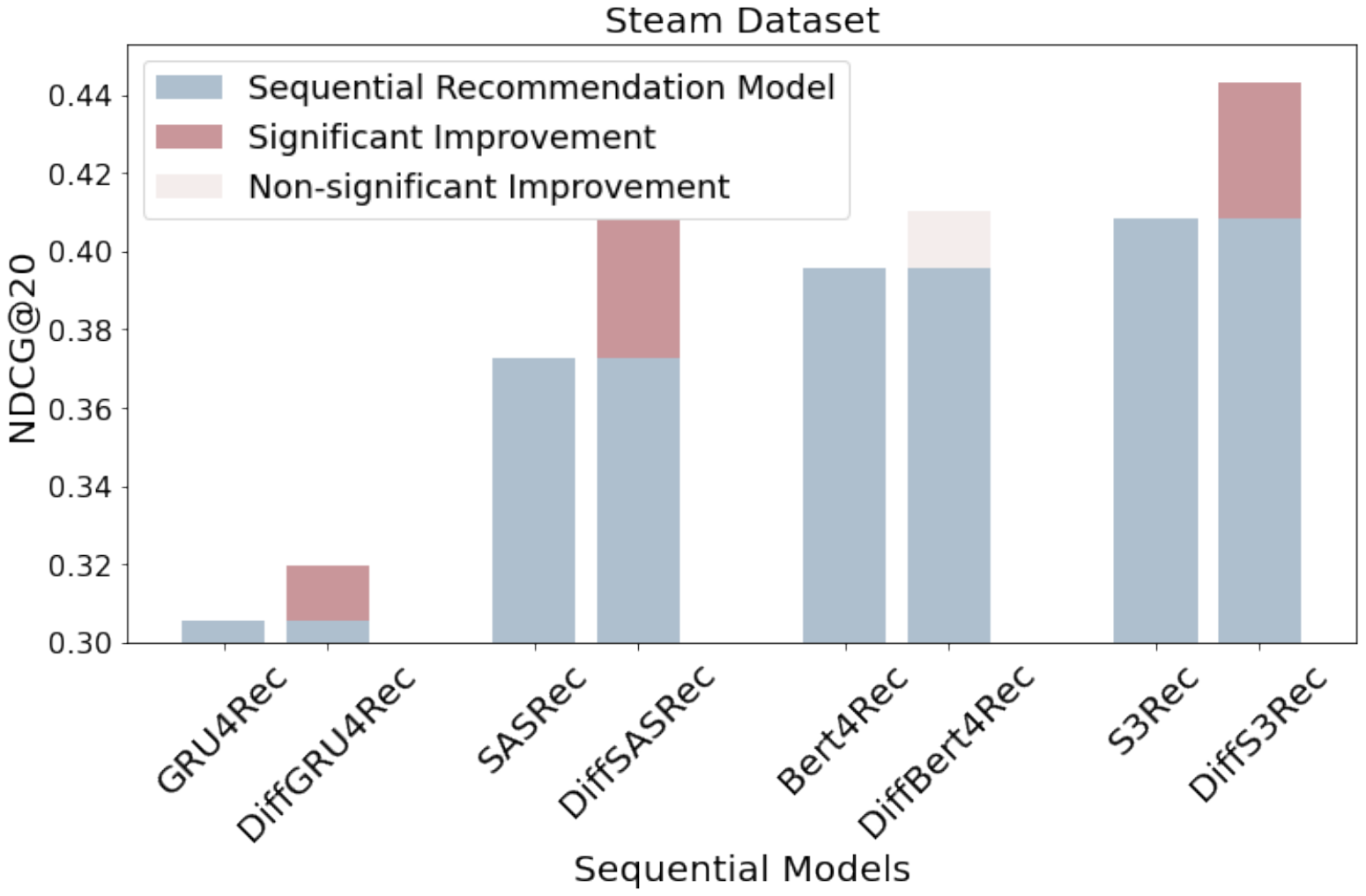}
    \end{subfigure}
    \caption{Recommendation performance of different KG and sequential recommendation models with or without our \zx{directional noise and the used linear transformer}. 
    A red colour on the bar indicates the significant difference between the tested models and their corresponding directional diffusion-enhanced approaches based on the paired t-test with $p<0.05$.}
\label{fig:gen}
\end{figure}

\vspace{+4mm}
\pageenlarge{2}
\noindent \textbf{Generalisation (RQ4):}
In this work, we \io{have introduced a} directional noise in the diffusion process \zx{(see Section~\ref{sec:directional})}. \io{We have also leveraged} a linear transformer to enable effective denoising within the diffusion process as a methodology \zx{(see Section~\ref{sec:diffgt})}. \io{We thoroughly investigated and evaluated such techniques in the top-k recommendation task using graph neural recommenders.}
In this section, we extend the use of \io{such techniques} to \io{other recommendation models so as to assess the generalisation of our proposed innovations beyond graph neural recommenders}. \io{First, we apply the techniques to} knowledge graph (KG) recommender\update{s} \io{on} the Amazon-Book\cite{mcauley2015image} and MIND\cite{wu2020mind} datasets. \io{Then, we investigate their effectiveness using}  sequential recommender\update{s} \io{on} the Beauty\cite{mcauley2015image} and Steam\cite{kang2018self} datasets.
Specifically, we integrate the diffusion process with \io{a} directional noise into the obtained user/item embeddings once \io{they have been} encoded \xw{by a} \io{knowledge graph (KG)} or \io{a} sequential \io{recommend\update{er}}.
Then, we denoise these noisy embeddings using the linear transformer. 
Figure~\ref{fig:gen} \io{presents} the \io{performance of several recent} KG and sequential \io{recommend\update{ers}}, when \io{we integrate} our designed directional noise and the linear transformer \io{techniques}.
\io{From} Figure~\ref{fig:gen}, we observe that the KG recommenders (CKE~\cite{zhang_collaborative_2016}, KGAT~\cite{wang_kgat_2019}, KGIN~\cite{wang_learning_2021}, KGCL~\cite{yang_knowledge_2022}, MCCLK~\cite{zou2022multi}) \io{augmented with} our diffusion methodology significantly outperform their original  \io{variants} in 90\% tested cases (9 out of 10 instances) on the Amazon-Book and MIND datasets. \io{Such} notable improvements \io{consolidate} the effectiveness of our \io{proposed} directional diffusion and linear transformer \io{techniques}, 
since the encoding methods (i.e., graph convolution) used in \io{the} KG recommenders are similar to those in \io{the} \zx{graph neural recommenders}.
\io{These results}  highlight the strong \xw{generalisation} of our \zx{injected directional noise and the used linear tranformer} \io{to} KG-enriched recommend\update{er}s.
\io{Next, we move to the sequential recommenders that aim to} predict the next items \xw{that \io{the} users \io{will interact with}}. 
\io{Figure~\ref{fig:gen} shows that}
the sequential recommenders (GRU4Rec~\cite{hidasi2015session}, SASRec~\cite{kang2018self}, BERT4Rec~\cite{sun2019bert4rec}, S3Rec~\cite{zhou2020s3}), \io{augmented with} our \io{new} diffusion \io{approach}, \zx{which includes both directional noise and a linear transformer},
show \io{a} superior performance \io{for} 4 out of 5 \io{models} (except Bert4Rec) on both the Beauty and Steam datasets, \io{compared to their original counter-parts}.
The absence of \io{a} significant improvement in Bert4Rec could be attributed to Bert4Rec's bi-directional mechanism, which is inherently more sensitive to noise due to its reliance on the full contextual understanding of the entire sequence.
\io{In this case, the} \io{injected} directional noise might disrupt this bi-directional balance, resulting in \io{a}  no notable \io{performance} improvement. 
In contrast, SASRec, which uses a unidirectional self-attention, \zx{appears to be} more robust to such a noise as it does not rely on the entire \xw{sequence of} interactions. \looseness -1

In response to RQ4, we conclude that our proposed directional noise and the used linear transformer can effectively generalise to both KG-enriched and sequential recommendation models.


\vspace{-2mm}
\pageenlarge{2}
\section{Conclusions}
In this \io{paper}, we introduced the Diffusion Graph Transformer (DiffGT), a novel \zx{diffusion} model for top-$k$ recommendation \zx{that enhances the \io{user/item} representations in order to effectively address the problem of \io{noise in the users'} implicit interactions \io{with the recommend\update{er} system}.}
\zx{Specifically, DiffGT \io{included} a novel graph transformer architecture to address the insufficient model architecture \io{used by} existing graph neural recommenders.}
\xw{\io{We showed} the anisotropic nature of recommendation data and \io{proposed} the use of directional noise}
\zx{to improve the diffusion process, ensuring \io{that} the used noise effectively aligns with the observed characteristics of recommendation data.}
\io{Our extensive experiments on three datasets showed that} DiffGT significantly outperformed ten strong existing baselines, \zx{including the \io{existing} out-of-box \io{state-of-the-art} DiffRec diffusion model, on the top-$k$ recommendation task.}
\zx{In \io{addition}, our ablation study showed that the incorporation of a directional noise \io{or} the linear transformer \io{in DiffGT}  resulted in a \io{significantly} improved performance.}
\zx{We also addressed the unconditioned diffusion paradigm \io{inherent to} the out-of-box diffusion model by using the average interacted embedding of the target user or item as a condition. Our \io{experimental} results showed that this \io{applied} condition can effectively guide the reverse \zx{process of the diffusion}.}
\zx{Moreover, we used side information to enrich the bipartite interaction graph and showed that using side information is a promising approach to further enhance the diffusion process \io{and improve the recommendation performance}.}
\zx{\io{For improving the efficiency \io{of DiffGT}}, we \io{proposed} a sampled denoising method and a linear transformer to efficiently enable the reverse process \io{within the diffusion} \io{without any} significant degradation in the recommendation performance.}
\zx{Finally, beyond graph neural recommenders, we showed that the effectiveness of our directional noise and the used linear transformer \io{effectively} generalised to other recommenders, such as \io{knowledge graph}-enriched and sequential recommenders.}

\balance
\bibliographystyle{ACM-Reference-Format}
\bibliography{reference}

\end{document}